\documentclass[superscriptaddress,pra,aps,twocolumn,10pt,longbibliography]{revtex4-1}

\usepackage{bm}
\usepackage{amsmath}
\usepackage{amssymb}
\usepackage{amsfonts}
\usepackage{graphicx}
\usepackage{float}
\usepackage{grffile}
\usepackage{natbib}
\usepackage{epstopdf}
\usepackage[pdftex]{color}
\usepackage[dvipsnames]{xcolor}
\usepackage[english]{babel}
\usepackage{hyperref}
\usepackage{verbatim}
\usepackage[normalem]{ulem}
\usepackage{float}
\usepackage[boxed]{algorithm}
\usepackage{algpseudocode}
\usepackage{soul}
\hypersetup{
    colorlinks=true,       % false: boxed links; true: colored links
    linkcolor=cyan,          % color of internal links
    citecolor=magenta,        % color of links to bibliography
    filecolor=magenta,      % color of file links
    urlcolor=cyan,           % color of external links
    runcolor=cyan
}

\newcommand{\ie}{{\it i.e. }}

\newcommand{\eg}{{\it e.g. }}

\def\eff{\mathrm{eff}}
\DeclareMathOperator{\tr}{Tr}

\newcommand{\iiprod}[3]{\langle #1 \rvert #2 \lvert #3 \rangle}

\newcommand{\avg}[1]{\left< #1 \right>}
\newcommand{\bra}[1]{\langle#1|}
\newcommand{\ket}[1]{|#1\rangle}

\def\config{\{\sigma_i\}}
\newcommand{\confign}[1]{\{\sigma_i^{(#1)}\}}
\def\add{\mathrm{add}}

\newcommand{\beq}{\begin{eqnarray}}
\newcommand{\eeq}{\end{eqnarray}}
\newcommand{\bes} {\begin{subequations}}
\newcommand{\ees} {\end{subequations}}

\newcommand{\ignore}[1]{}

\graphicspath{{figs/}}

\begin{document}

\title{Perils of Embedding for Quantum Sampling}

\author{Jeffrey Marshall}
\affiliation{QuAIL, NASA Ames Research Center, Moffett Field, CA 94035, USA}
\affiliation{USRA Research Institute for Advanced Computer Science, Mountain View, CA 94043, USA}

\author{Gianni Mossi}
\affiliation{QuAIL, NASA Ames Research Center, Moffett Field, CA 94035, USA}
\affiliation{KBR, 601 Jefferson St., Houston, TX 77002, USA}

\author{Eleanor G. Rieffel}
\affiliation{QuAIL, NASA Ames Research Center, Moffett Field, CA 94035, USA}

\begin{abstract}
Given quantum hardware that enables sampling from a family of natively implemented Hamiltonians, how well can one use that hardware to sample from a Hamiltonian outside that family? A common approach is to minor embed the desired Hamiltonian in a native Hamiltonian.
In \href{https://link.aps.org/doi/10.1103/PhysRevResearch.2.023020}{Phys. Rev. Research \textbf{2}, 023020 (2020)} \cite{perilsOfEmbedding_PRR} it was shown that minor embedding
can be detrimental for classical thermal sampling. Here, we generalize these results by considering quantum thermal sampling in the transverse-field Ising model, \ie sampling a Hamiltonian with non-zero off diagonal terms. 
In the quantum case, loosely speaking, it is even harder to preserve the correct distribution properties, since the local transverse fields affect the physical qubits in the embedding in a manner that cannot be lifted by setting an appropriate energy scale, as in the classical case.
To study these systems numerically we introduce a modification to standard cluster update quantum Monte-Carlo (QMC) techniques, which allows us to much more efficiently obtain thermal samples of an embedded Hamiltonian, enabling us
to simulate systems of much larger sizes and larger transverse-field strengths than would otherwise be possible.
Our numerics focus on models that can be implemented on current quantum devices using planar two-dimensional lattices, which exhibit a phase transition driven by the transverse field strength.
Our results include:
i) An estimate on the probability to sample the logical subspace directly as a function of transverse-field, temperature, and total system size, which agrees with QMC simulations. 
ii) We show that typically measured observables (diagonal energy and magnetization) are biased by the embedding process, in the regime of intermediate transverse-field strength, meaning that the extracted values are not the same as in the native model.
iii) By considering individual embedding realizations akin to `realizations of disorder', we provide numerical evidence suggesting that as the embedding size is increased,
the critical point shifts to increasingly large values of the transverse-field.
\end{abstract}
\maketitle

\section{Introduction}

The last several years have seen the emergence of a variety of quantum processors. While the progress has been rapid, engineering constraints limit which Hamiltonians can be natively implemented on this hardware. 
One approach, the gate-model approach, is to break down the desired computation into gates with Hamiltonians acting on only a small number of qubits.
Global approaches where Hamiltonians act on all qubits simultaneously, including analog systems such as quantum annealing \cite{Apolloni:1989qy,kadowaki_quantum_1998} and population transfer \cite{pop-transfer}, and those more digitally oriented such as global pulsing \cite{blueprint-suprem} and specialized simulation \cite{digital-analog,analogue-otoc}, enable quantum computations that cannot be effectively carried out on current devices in the gate-model. As such, these global paradigms will continue to complement what can be done in the gate-model for years to come.

In these global approaches however, restricted topologies and couplings available limit the set of natively implementable Hamiltonians. This leads to a general question:
Given a device that can implement a set of Hamiltonians $\cal H$, how well can it be used to study a Hamiltonian $H \notin \cal H$? This question applies to both time-dependent and time-independent Hamiltonians. Here, we concentrate on one possible technique, minor embedding \cite{Choi1}, that is commonly used in quantum annealing and can be applied in other settings as well. 
Within this framework, we consider the  general case of thermal sampling from embedded quantum Hamiltonians, extending \cite{perilsOfEmbedding_PRR} that considered classical Hamiltonians.

Thermal sampling of the Gibbs distribution induced by a Hamiltonian, $H=H^\dag$,  at temperature $T=1/\beta$, involves generating samples in some basis $\{|z\rangle \}_z$ with probability
\begin{equation}
    P_z = \frac{1}{Z}\langle z| e^{-\beta H}|z\rangle,
\end{equation}
where $Z=\mathrm{Tr}[e^{-\beta H}]$ is the partition function. In practice this is done either experimentally or by numerical simulations, but in either case technical or computational considerations usually limit the choice of the basis $\{|z\rangle \}_z$.

The case where $\{|z\rangle \}_z$ is the eigenbasis of $H$ is particularly simple since then $P_z = \frac{1}{Z}e^{-\beta E_z}$, where $H|z\rangle = E_z|z\rangle$ (\ie $E_z$ is the energy of the eigenstate $\lvert z \rangle$). In this case the relative sampling probability depends only on the difference in the energy levels,
\begin{equation}
\label{eq:transition}
    \frac{P_{z_1}}{P_{z_2}} = e^{-\beta (E_{z_1}-E_{z_2})}.
\end{equation}
This is most often the case for classical thermal systems, where many very successful thermal sampling algorithms have been developed over the years. Particularly significant among these are simulated annealing and parallel tempering (replica exchange Monte-Carlo), where transitions between states are governed by Eq.~\eqref{eq:transition}.

Conversely, it often happens that for quantum mechanical systems $H$ is not diagonal in the physically implementable $\{|z\rangle \}_z$ basis, and for this reason we will refer to such a case as quantum thermal sampling. There are additional challenges in sampling from such a system using conventional computational methods. In fact, in cases which are not amenable to quantum Monte-Carlo (QMC) \eg due to a sign problem (non-stoquastic), there are generally no known methods available to efficiently sample the Gibbs distribution.

Sampling from thermal distributions, whether classical or quantum, has many applications in science. Thermal quantities can be used to track fundamental changes in the symmetries of a system, \ie as a phase transition occurs. Efficiently performing such a sampling allows one to therefore probe criticality, which has use in material science, and physics theory. Another relevant example is to solve optimization problems, which can be achieved by thermal sampling at low enough temperatures, where the problem is encoded in the spectrum of the Hamiltonian.
In addition, classical and quantum thermal sampling has application in the field of machine learning \cite{boltzmann-machine, Amin:boltzmann}.

Emerging technology, such as quantum annealers, have been proposed, and used, to sample from such distributions, for both classical \cite{adachi,perdomo,DW2x-gibbs-ML,QAAAN,boltzmann-li-qac,boltzmann-galaxy, dw-shastry-sutherland} and quantum \cite{qvae,harris-tfim,king-topological,Amin:boltzmann,izquierdo2020testing, Griffiths-McCoy, king-top-2021} thermal sampling tasks.  This is a promising avenue of experimental exploration since classical numerical implementations can often struggle to sample with sufficient accuracy, in particular at low enough temperatures \cite{hastings-obstructions}.

As mentioned previously, limitations in the topology restrict which Hamiltonians can be natively implemented. Minor embedding \cite{Choi1} is a technique frequently used in the quantum annealing community to effectively increase the connectivity, by introducing additional variables. Assuming the newly introduced parameters are chosen appropriately, one can guarantee the low lying energy spectrum of the embedded \textit{classical} Hamiltonian matches the desired one. For optimization purposes, embedding can therefore be a powerful tool to circumvent restrictions imposed by a physical device.

In fact, since the \textit{entire} spectrum of a native classical Hamiltonian is preserved by the embedding process (energy levels only have a constant shift), classical thermal sampling can, in principle, be performed on the embedded (classical) Hamiltonian. (Although in practice there are additional complications as discussed in  Ref.~\cite{perilsOfEmbedding_PRR}.)
The same is not true in the quantum case however. In particular, there are no guarantees, nor can one set parameters appropriately, such that the spectrum of the desired (native) \textit{quantum} Hamiltonian is preserved by the embedding.

Since this type of quantum sampling is increasingly an area of focus (\eg for use in physics simulations), it is important to have a better understanding of the implications of embedding in terms of how it affects sampling properties.
Prior works (such as \cite{harris-tfim,dw-shastry-sutherland,logical-qubit-dwave, dw-fast-clique}) have addressed possible issues due to embedding bias in various manners, from searching for more efficient (\ie smaller) embeddings, performing post-processing of samples, and tuning parameters in such a way that the desired system is more faithfully reproduced.
Moreover, in any analog device with limited connectivity, embedding issues can similarly arise. Possible examples include certain quantum chemistry simulations (such as real-time evolution of Fermi and Bose-Hubbard models \cite{fermi-hubbard-1d,analogue-otoc}), or population transfer techniques \cite{pop-transfer}.

Here we study the effect of embedding on quantum thermal sampling of the TFIM, using a physically relevant two-dimensional system (with ordered couplings) which can be implemented easily on current and near-term hardware, both with and without embedding. 
This model therefore can serve as a test bed for studying our results experimentally.
We take a broad view of the problem, and aim to shed light on the effect a fixed embedding has on a system by separating it from other possible sources of noise and distortion, \ie we will not assume the ability to freely tune parameters or post-process samples.

We perform our analysis by introducing a modification of the standard quantum Monte-Carlo (QMC) algorithm applied to the TFIM, which otherwise (as we will discuss below), becomes extremely inefficient once embedding is introduced.
We believe this will be very helpful in studying the effect of embedding in physically implementable models on current devices with many free parameters, and thus aid in the setting of such parameters. Though our scheme is static (sampling occurs at fixed parameter values), it is trivial to modify it to work in a dynamic setting, such as quantum annealing.
In this work, to isolate the effects of the embedding, we focus on a rejection-based sampling scheme, where we reject any sample which is not in the logical subspace (the subspace of configurations which are well defined in the native model).

Our contributions include
\begin{itemize}
    \item An estimation of the probability for sampling the logical subspace directly as a function of system size, embedding size, and transverse field (Sect. \ref{sect:PL})
    \item An introduction of a QMC scheme which can more efficiently sample the logical subspace of an embedded Hamiltonian, as compared to the naive approach of performing QMC on the full embedded Hamiltonian which results in many wasted samples (those coming from outside the logical subspace) (Sect.~\ref{sect:qmc-main}, Appendix \ref{sect:appendix-qmc})
    \item Observations of measured observables being distorted by embedding. By considering embedding realizations on a similar footing to disorder realizations in statistical physics, we show a linear (in the embedding size) shifting of the critical point to larger values of transverse-field (Sects. \ref{sect:sampling-bias}, \ref{sec:scaling_properties}).
\end{itemize}

\section{Methods}

\subsection{Problem Studied \label{sect:problem}}

We study as our `native' problem an anti-ferromagnetic transverse-field Ising model (TFIM)  on a square lattice with side-length $L$ (with $N=L^2$ qubits) defined by Hamiltonian
\begin{equation}
\label{eq:hamiltonian}
\begin{split}
    & H=\Gamma H_\Gamma + \Delta H_\Delta \\
     & := -\Gamma \sum_{i=1}^N \sigma_i^x + \Delta \sum_{\langle i,j\rangle } \sigma_i^z \sigma_j^z,
    \end{split}
\end{equation}
where the angle brackets indicate the sum is only over neighbours on the square lattice (with free boundary conditions).
Here $\Gamma, \Delta \ge 0$.

This model exhibits a phase transition in the transverse-field at non-zero temperature, for which analytical and numerical results exists \cite{Elliott_1971, 2d-qmc}. The $T=0$ quantum critical point is documented at $\Gamma/\Delta = 3.044$ \cite{2d-qmc}.

\subsection{Embedding Details \label{sect:embedding}}

Here we describe how we embed a Hamiltonian of the form Eq.~\eqref{eq:hamiltonian} (or in general any TFIM Hamiltonian) to a physical hardware graph (the hardware graph we ultimately use in simulations will effectively be a random one, which we will discuss below). A useful reference with more information about embedding in general is Ref.~\cite{Choi1}.

If a given hardware graph does not respect the topology of the desired Hamiltonian (in our case Eq.~\eqref{eq:hamiltonian}), one must embed the `native' graph to the hardware.
To do this, each qubit in the native model is constructed from several physical qubits, coupled ferromagnetically as a linear chain in the $z$-direction, with equal strength $J_F<0$. We call these `logical qubits'. In our simulations we fix $J_F=-2$ in units of $\Delta$ (which is often the largest value used in experiments \cite{perilsOfEmbedding_PRR}). 

For logical qubit $l$ composed of $n_l$ physical qubits which can each be indexed as $l_i$, where $l=1,\dots N$, and $i=1,\dots n_l$, the Hamiltonian defining a logical qubit $l$ is
\begin{equation}\label{eq:Hl}
    \tilde{H}^{(l)} = J_F \sum_{{i}=1}^{n_l-1 }\sigma_{l_i}^z \sigma_{{l_i}+1}^z.
\end{equation}
With this, the full embedded Hamiltonian, of $\tilde{N}= \sum_{l=1}^N n_l$ physical qubits, is 
\begin{equation}
\label{eq:H_embed}
    \tilde{H} = \Gamma \tilde{H}_\Gamma + \Delta \sum_{l=1}^N \tilde{H}^{(l)} + \Delta \tilde{H}_{\Delta}
\end{equation}
where $\tilde{H}_\Delta$ couples the logical qubits through physical qubits connected in the hardware graph.
In particular, if an embedding exists, it guarantees for any two logical qubits coupled in the native graph, there are physical qubits of each logical qubit that can be coupled in the hardware graph. 
Then we can write
\begin{equation}
\label{eq:Hj}
    \tilde{H}_\Delta = \sum_{\langle l, k \rangle} \sigma_{{c(l, k)}}^z\sigma_{{c(k, l)}}^z,
\end{equation}
where we have introduced the function $c$, such that $c(l, k) \in 1, \dots , n_l$ returns a qubit index for a qubit in the logical qubit $l$, which has a coupling to the physical qubit $c(k, l) \in 1, \dots, n_k$, contained in logical qubit $k$ (which exists assuming the embedding exists). That is, $c$ provides the physical qubit mapping between two logical qubits.
The angle brackets denote the indices only run over those defined in the problem (here a square lattice).

Note that the global transverse-field induces the Pauli $\sigma^x$ operator on each physical qubit individually; $\tilde{H}_\Gamma = -\sum_{i=1}^{\tilde{N}} \sigma_i^x$.

The result of the above, is that now the embedded Hamiltonian respects the topology of the device.
Moreover, in the case $\Gamma=0$, for large enough $|J_F|$ the ground subspace of $\tilde{H}$ is the same as the original Hamiltonian $H$ \cite{perilsOfEmbedding_PRR}.

We use $K\ge 1$ to refer to the average embedding size, \ie $K=\tilde{N}/N$.
In our simulations, for a given logical problem of $N$ variables, and a specified embedding size $K\ge 1$ (which need not be integer), we find the closest integer $\tilde{N}$ such that $\frac{\tilde{N}}{N} \approx K$. 
The number of additional variables introduced is $D=\tilde{N}-N$, which are randomly distributed as chains over the $N$ logical variables. 
For example, if $D=1$, there will be one non-trivial chain (logical qubit) of size 2, but if $D=2$, this could be either two chains of size 2, or one of size 3.
For integer $K$ we pick each chain to be of size $K$.
For non-integer cases where $K < 3$, we consider a distribution of chain sizes from  $\{1, 2, 3\}$. 

\subsubsection{Random Graph Embedding \label{sect:physical_graph}}

Our main model of embedding is intended to represent a somewhat realistic model of embedding, whilst at the same time not being restricted to any particular topology.
We do this by coupling logical spins by a single physical bond, randomly selected. 
In particular, once the distribution of the logical qubit sizes is fixed as above, if any two logical qubits need to be coupled (\ie they coupled in the native graph), we pick a random pair of physical qubits between the logical qubits to couple. 
This is intended to model the commonly-encountered situation where one needs to connect two physical spins but does not have the freedom to arbitrarily choose the point of connection between their respective ferromagnetic chains, without explicitly introducing any specific assumptions on the hardware's topology that forces this constraint.

\subsection{Quantum Monte-Carlo \label{sect:qmc-main}}
In order to estimate quantum thermal properties for problems of sizes which are beyond exact numerics we implement a QMC sampler with Wolff cluster updates in the imaginary time direction only. 
For a clear and concise description of QMC for the transverse-field Ising model, we point the reader to Appendix A in Ref.~\cite{qmc_qa}.
The clusters are built as in Ref. \cite{2d-qmc}, although we do not extend them in the real direction (which becomes inefficient for certain problems). Instead, we flip each cluster according to the (spatial) Metropolis probability, as in Ref.~\cite{boixo:14}.

In particular, at a high-level, the QMC cluster update procedure we use proceeds as follows:
\begin{enumerate}
    \item Set-up: Pick the number of imaginary time steps (`replicas'), $\ell$, and a random initial configuration of the total $N\times \ell$ spins (where $N$ is the system size, repeated for each imaginary time).
    \item For each imaginary world-line of $\ell$ spins, we group it into clusters. A cluster is a set of neighbouring spins aligned identically, with each spin joined to the cluster with probability $1-\exp(-2\beta_{eff} J^\perp)$ where $\beta_{eff}=\beta/\ell$, $J^\perp = -\frac{1}{2\beta_{eff}} \log \tanh \beta_{eff} \Gamma$.
    \item Flip all of the spins in a cluster with the standard Metropolis acceptance probability $p=\min\{1, e^{-\beta \Delta E}\}$, where the energy change $\Delta E$ is computed in the real (spatial) direction only.
\end{enumerate}
Steps two and three are repeated as desired.

We always run $10^3$ thermalization sweeps before taking statistics (lowering the temperature incrementally to the target). The choice of the number of time-slices and samples is discussed in Appendix \ref{sect:sim_params}.

Our QMC code runs in two modes.
\subsubsection{Rejection based QMC \label{sect:rejection-qmc}}
In \textit{rejection mode}, we implement the standard QMC algorithm outlined above, but we only takes statistics from a time-slice if it is a logical configuration. This mode is used in order to estimate the probability of obtaining a logical configuration, as would occur in a physical device, which can be very inefficient depending on the parameters used. We discuss this in more detail in Sect.~\ref{sect:PL}.

\subsubsection{Logically-constrained QMC}
\begin{figure}
    \centering
    \includegraphics[width=0.98\columnwidth]{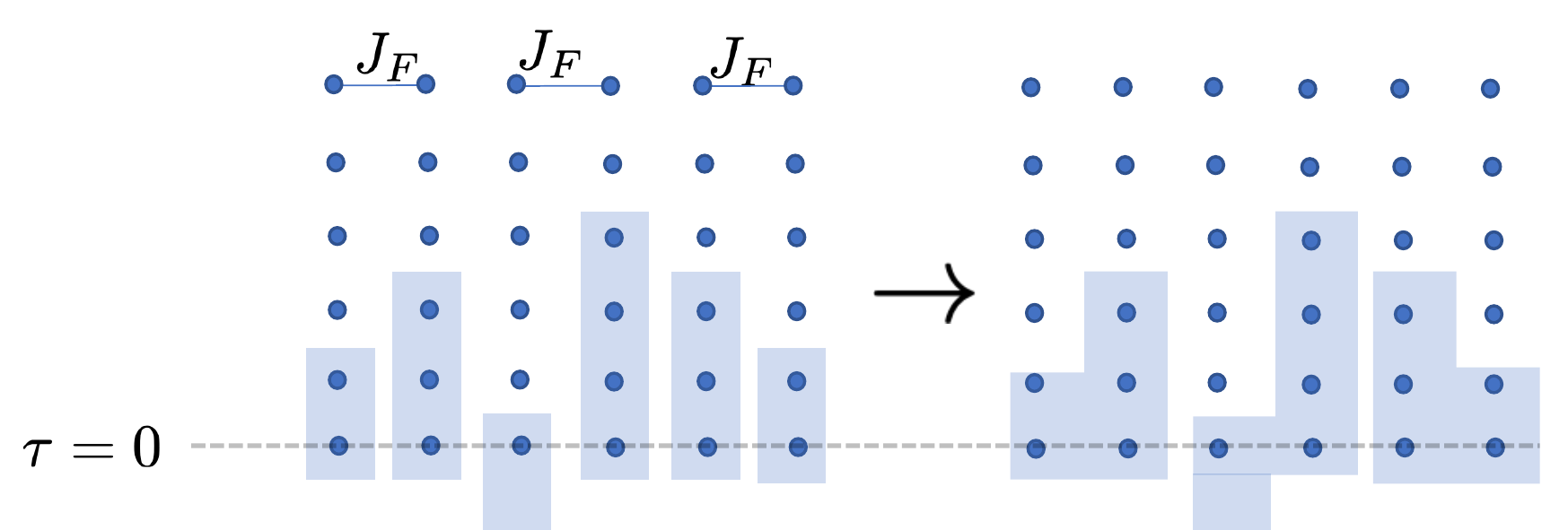}
    \caption{\textbf{Cartoon of update scheme for LC-QMC.} Shown is a schematic of the classical spins in QMC resulting from a 3 qubit native problem, where each qubit is embedded to size 2 with ferromagnetic strength $J_F$ (vertical direction is the imaginary-time $\tau$ axis, of which we only show a few time-slices). On the left, clusters are built as in standard imaginary-time cluster update QMC (we only highlight those going through $\tau=0$). On the right we join clusters traversing the $\tau=0$ time-slice, if they belong to the same logical variable. These are then flipped in accordance to the standard (spatial) Metropolis probability. This ensures the $\tau=0$ time-slice is constrained within the logical subspace (assuming it is initialized there), whilst preserving the detailed balance conditions of the QMC. There are periodic boundary conditions in the imaginary time direction (\eg the cluster of the third spin loops around). We outline the recipe in more detail in Appendix \ref{sect:appendix-qmc}.}
    \label{fig:lc-qmc}
\end{figure}

We introduce a modification to the standard QMC outline above, for use when embedding, which allows us to sample much larger system sizes than the rejection based code above. We call this, \textit{logically-constrained QMC} (LC-QMC). In this mode, we constrain the 0'th time-slice to always be a logical one, which means this slice can always be used for taking a measurement. To start, the initial state is random as in step 1 above, but the 0'th slice must be a logical configuration. Then, after building the clusters (as in step 2 above), we join all clusters through time-slice $\tau=0$ corresponding to a logical qubit, as demonstrated in Fig.~\ref{fig:lc-qmc}. A cluster is then flipped via the spatial Metropolis probability as in step 3 as usual. Due to how the clusters are constructed, the 0'th time-slice never leaves the logical subspace. In Appendix \ref{sect:appendix-qmc} we outline the calculation to show that the detailed balance condition holds for this update, and that therefore the statistics of this method are identical to that of the above. We also plot in Fig.~\ref{fig:constrained_qmc} a comparison of the two methods for a small system, showing that they agree with an exact computation.

We lastly mention that this technique can be trivially implemented in the context of Simulated Quantum Annealing (SQA) \cite{SQA-original, Crosson-SQA}, and as such can more faithfully model the embedding effect during a quantum evolution.

%%%%%%%%%%
\section{Results}
We wish to understand to what extent samples taken in the $z$ eigen-basis from the ideal distribution $\exp(-\beta H)$ can be obtained by sampling instead from $\tilde{H}$, the embedded Hamiltonian:  $\exp(-\beta \tilde{H})$, when $\Gamma>0$. 
We will focus on the task of direct sampling, where any sample not from the logical subspace is discarded.

In practice unembedding schemes can also be used, \ie postprocessing the samples to project to the logical subspace. These techniques will generally alter the distribution of samples however, and can therefore obscure the physics due to the embedding itself. Our goal in this work is to isolate the effects of embedding, and therefore we consider only the rejection based sampling described above, and leave the analysis of unembedding as a future task.

\subsection{The sampling problem I: direct sampling \label{sect:PL}}
In the scenario where we are interested in only measuring logical configurations (discarding any configuration with broken chains), one key quantity of interest is the probability to observe a sample from the logical subspace when making measurements in the computational basis, which we denote $P_L$. This is relevant for experimental realizations of such thermal embedded sampling, and is computed via
\begin{equation}
\label{eq:PL}
    P_L = \frac{1}{\tilde{Z}}\sum_{z_L} \langle z_L| e^{-\beta  \tilde{H}}|z_L\rangle
\end{equation}
where the sum is over logical configurations $z_L$, $\tilde{H}$ is the embedded Hamiltonian (as in Eq.~\eqref{eq:H_embed}), and the partition function is $\tilde{Z} = \mathrm{Tr}e^{-\beta  \tilde{H}}$.

In the worst case, it is easy to see the probability to sample the logical subspace is exponentially small in the \textit{total} system size. This can be seen by considering the regime where the transverse-field dominates, $\Gamma \gg \Delta$, with the probability of observing a logical configuration is $P_L \approx 2^{-N(K-1)}$ (with equality in the case $\Delta=0$). This is found by computing
\begin{equation}
  e^{\beta \Gamma \sum_{i=1}^{NK} \sigma_i^x}= \prod_{i=1}^{NK}  e^{\beta \Gamma \sigma_i^x} = \prod_{i=1}^{NK} (\cosh \beta \Gamma + \sigma_i^x \sinh \beta \Gamma)
\end{equation}
giving in the case $\Delta=0$, partition function $\tilde{Z}=2^{NK}(\cosh \beta \Gamma)^{NK}$. The probability of observing a logical configuration is therefore $2^N / 2^{NK}$,
using that sum in Eq.~\eqref{eq:PL} is over $2^N$ terms, and $\langle z_L| \sigma_i^x|z_L\rangle=0$.

For arbitrary $\Gamma/\Delta$ we can approximate $P_L$ as follows. First, consider the local Hamiltonian for a single chain embedding of size two,
\begin{equation}
\label{eq:2chain}
    H = -\Gamma(\sigma_1^x + \sigma_2^x) + \Delta J_F \sigma_1^z \sigma_2^z.
\end{equation}
From this we can compute (via Eq.~\eqref{eq:PL}), defining $E:=\sqrt{4 \Gamma ^2+\Delta^2 J_F^2}$,
\begin{equation}
\label{eq:PL_2chain}
    p_L = \frac{\Delta |J_F| \sinh \left(\beta E\right)+E \cosh \left(\beta  E\right)+E e^{\beta \Delta |J_F|}}{2E \left(\cosh \left(\beta  E\right)+\cosh (\beta  \Delta |J_F|)\right)},
\end{equation}
where we use lower case $p_L$ to indicate this is the logical probability for this reduced system.

To extend this analysis beyond two qubits, we assume the breaking of chains is statistically independent (which holds in the regime where $|J_F|$ is sufficiently large compared to the problem couplings). With this, we make the Ansatz that the total logical subspace sampling probability follows $P_L \approx p_l^{N(K-1)}$.
Note, this equation gives the same limit as found in Ref.~\cite{perilsOfEmbedding_PRR} for zero transverse-field. This formula also obtains the $\Delta=0$ limit discussed above.

\begin{figure}
    \centering
    \includegraphics[width=1\columnwidth]{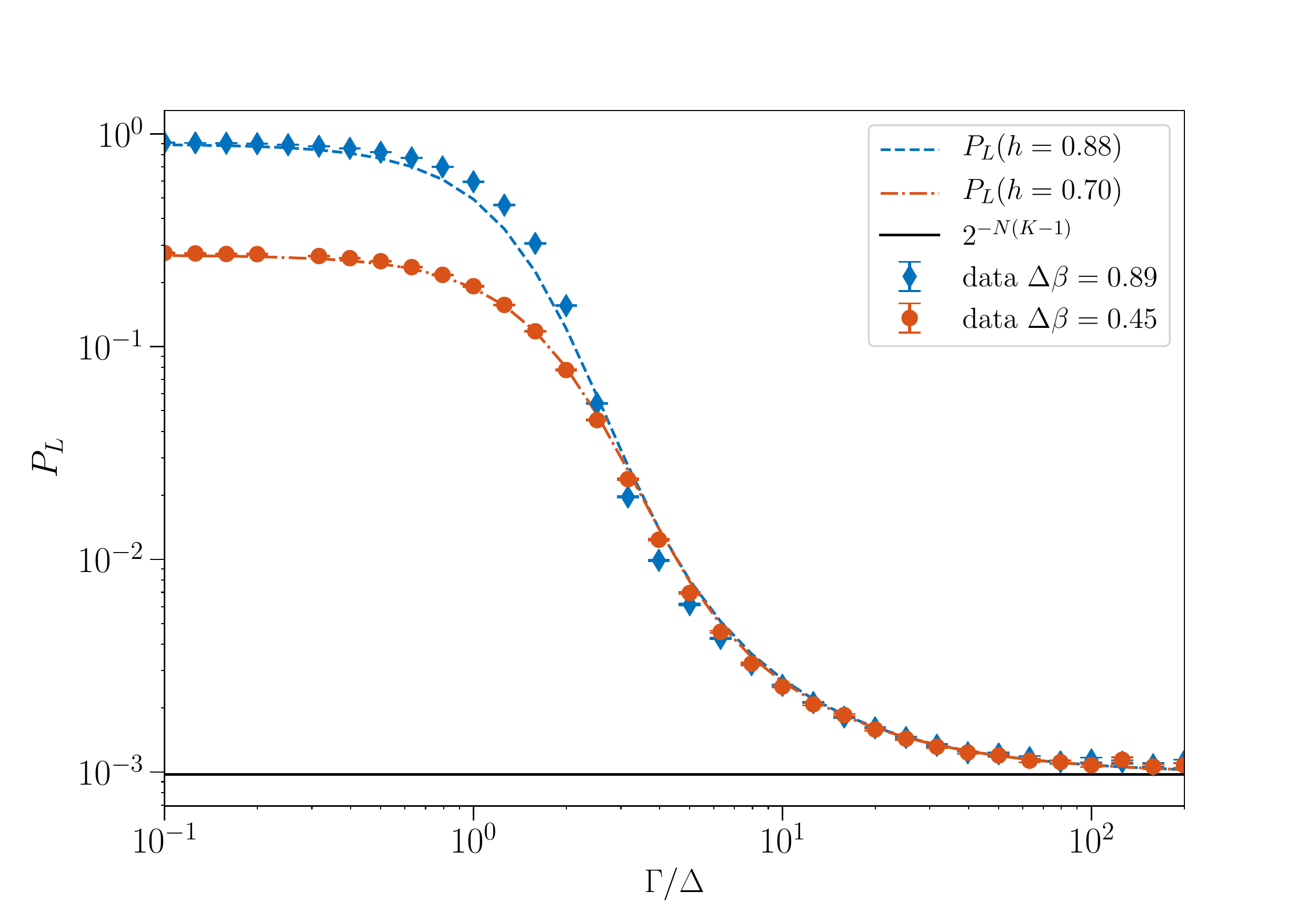}
    \caption{\textbf{Logical probability as a function of transverse-field}. For two temperatures we show data from our QMC simulations (dots/diamonds) overlayed with a curve fit as described in the main text, using local field $h$ as a fitting parameter (see legend). The problem studied here is a 2D anti-ferromagnet with side-length $L=10$ (100 qubits), and embedding $K=1.1$. The horizontal black solid line is the $\Gamma\rightarrow\infty$ result. Here we take at least $2^{17}$ samples per data point (error bars are the standard error as discussed in Appendix \ref{sect:sim_params}). To perform this simulation, we use the `rejection-based' QMC (Sect.~\ref{sect:rejection-qmc}).}
    \label{fig:PL}
\end{figure}

In order to account for the effect of the native problem Hamiltonian (the $J_{ij}$) on $p_L$, we can additionally include a ``mean field'' term to Eq.~\eqref{eq:2chain}, $H\rightarrow H + h(\sigma_1^z + \sigma_2^z)$, for which we can use $h$ as a fitting parameter. This parameter can then act to capture specific properties of the system.
An example of this curve fitting is shown in Fig.~\ref{fig:PL} for a relatively small embedding $K=1.1$, which captures the shape accurately for all $\Gamma$.

For observing phase transitions in such a model requires to probe transverse fields in the regime $\Gamma \in [1,10]$ (units of $\Delta$).
It is clear that for large enough problem sizes and embeddings, eventually it will be infeasible to directly sample the logical subspace. 
In order to observe $M$ logical samples, requires $O(M/P_L)$ total samples, which is growing exponentially in the parameters $N,K$.
This means the majority of all samples observed will contain broken logical spins, especially in the regime $\Gamma/\Delta \gtrsim 1$. 

This exponential reduction in sampling the logical subspace is also the reason why the standard rejection based QMC is not an  appropriate tool for study here, and motivates our introduction of the above described LC-QMC. We use this in the remainder of our analysis below.

\subsection{The sampling problem II: inherent bias}
\label{sect:sampling-bias}
In the case of zero transverse-field $\Gamma=0$, embedding only causes an overall energy shift to the logical subspace. Within the logical subspace, the difference between any two logical energy levels therefore remains unchanged, and if one directly samples the logical subspace (by discarding any illogical solution), the distribution will still be Boltzmann, by Eq.~\eqref{eq:transition}. This is explained in more detail in Ref~\cite{perilsOfEmbedding_PRR}.
This is not true for non-zero transverse-field; for $\Gamma>0$, directly sampling the logical subspace, by rejecting illogical configurations, inevitably introduces a bias to the statistics. 

\begin{figure}
    \centering
    \includegraphics[width=0.98\columnwidth]{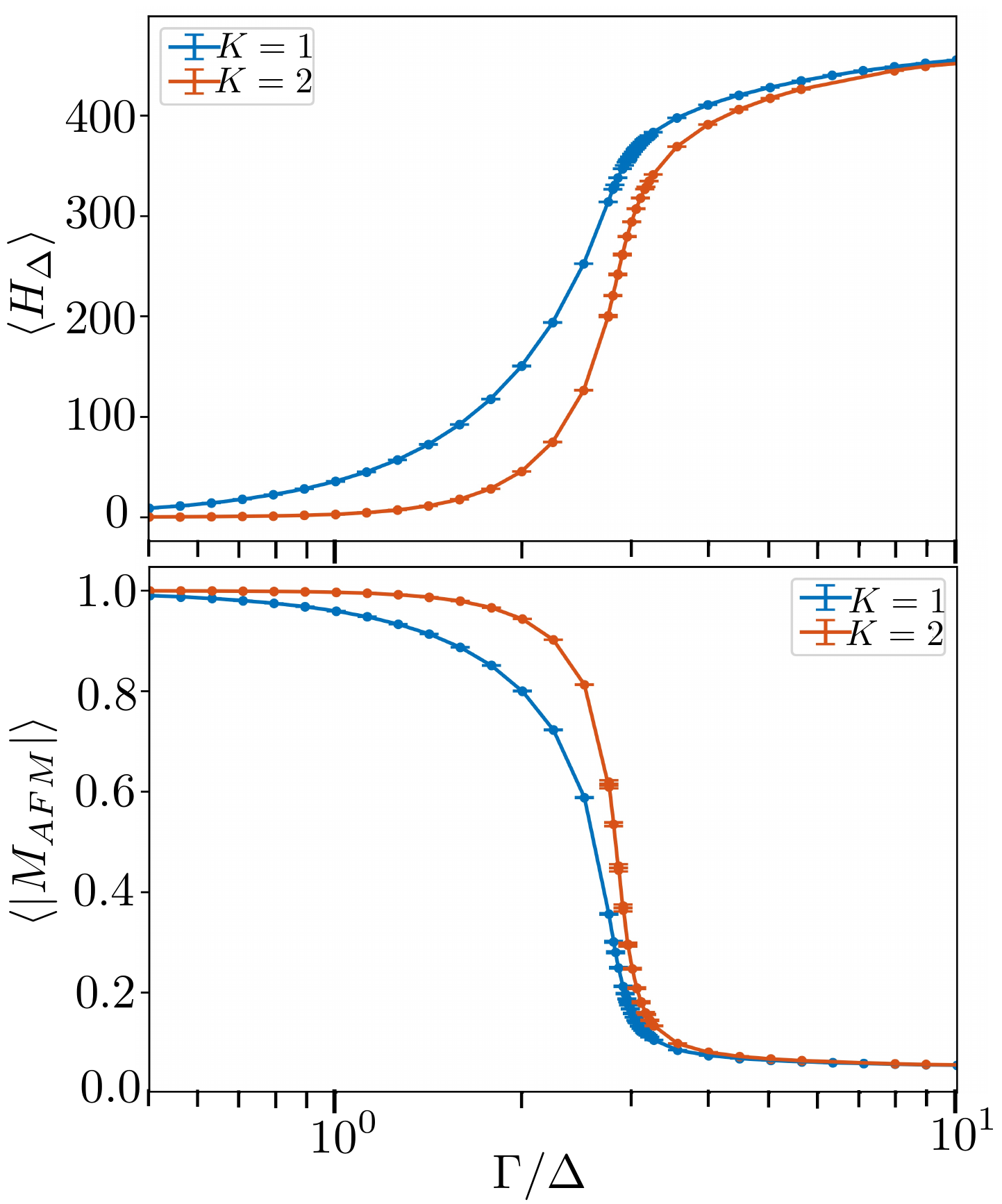}
    \caption{\textbf{Bias in observable values by embedding.}
    (Top) Expectation value of energy $\langle H_\Delta \rangle$ for native 2D lattice ($K=1$, top curve), and when embedding with $K=2$ (bottom curve). We have performed a shift so that the ground state has $E=0$.
    (Bottom) Expectation value of anti-ferromagnetic order parameter (absolute value), for native 2D lattice ($K=1$, bottom curve), and when embedding with $K=2$ (top curve).
    For both plots the lattice size is $L=16$, with $\Delta \beta = 1.645$ \cite{temp-note}. Here error bars are the standard error over independent samples.}
    \label{fig:Ep_M}
\end{figure}

Let us consider, for non-zero $\Gamma$, the probability to observe a particular logical configuration $z$. 
Let's call the eigenbasis of $H$ $(\tilde{H})$ as $|E_i\rangle$ ($|\tilde{E}_i\rangle$) with eigenvalues $E_i$ ($\tilde{E}_i$). 
If we write
\begin{equation}
    |E_i\rangle = \sum_z c^{(i)}_z |z\rangle,
\end{equation}
then
\begin{equation}
    P_z = \frac{1}{Z} \sum_{i=1}^{2^N} e^{-\beta E_i} |c_z^{(i)}|^2.
\end{equation}
We denote the equivalent probability by sampling the logical subspace of the embedded Hamiltonian $\tilde{H}$, with appropriate normalization, as
\begin{equation}
    \tilde{P}_z = \frac{1}{P_L}\frac{1}{\tilde{Z}} \sum_{i=1}^{2^{NK}} e^{-\beta \tilde{E}_i} |\tilde{c}_z^{(i)}|^2.
\end{equation}
If for any logical $z$ one has $\tilde{P}_z \neq P_z$, we say the sampling is biased by embedding. In the (`classical') case $\Gamma=0$ we have $\tilde{P}_z = P_z, \forall z$ since here the embedding simply shifts the logical spectrum. This is similarly true in the case $\Gamma/\Delta \rightarrow \infty$, as the distribution (over $\{z\}$) tends to the trivial uniform (or infinite temperature) distribution. In between these regimes however, the embedding will typically distort the distribution.

A bias here will typically result in the bias of any observable one wishes to measure, for example magnetization or the energy. 
For a diagonal (in $z$) logical observable $O=\sum_z O_z |z\rangle \langle z|$ one has
\begin{equation}
    \langle O\rangle = \frac{1}{{Z}}\mathrm{Tr} [O e^{-\beta {H}}] = \sum_z O_z P_z.
\end{equation}
It is clear that in general if sampling the embedded Hamiltonian instead, and computing $\langle O\rangle$ from the logical samples received, that if $P_z\neq \tilde{P}_z$, there is no guarantee one will compute the correct expectation value $\sum_z O_z P_z \neq \sum_z O_z \tilde{P}_z$.

In our system, the relevant order parameter is the anti-ferromagnetic (\ie staggered) magnetization $M_{AFM} = \frac{1}{N} \sum_{i=1}^N (-1)^{x_i+y_i} s_i$, where a spin configuration is given by $\mathbf{s}=(s_1,\dots, s_N)$, and each site $i$ has coordinates $(x_i,y_i)$. In Fig.~\ref{fig:Ep_M} for a system size $L=16$, we see that in general the embedded model does not compute the correct value for either the (diagonal) energy, nor the order parameter.

As expected, for small and large values of $\Gamma$ there is no bias (the former tending to the classical unbiased case, and the latter to the trivial infinite temperature case), but for intermediate $1\lesssim \Gamma/\Delta \lesssim 10$ there can be a significant deviation in the computed quantities. For this problem, the embedding causes a lower energy than expected and a higher magnetization (it is, in a certain sense, effectively lowering the temperature of the distribution).

\begin{figure}
    \centering
    \includegraphics[width=0.98\columnwidth]{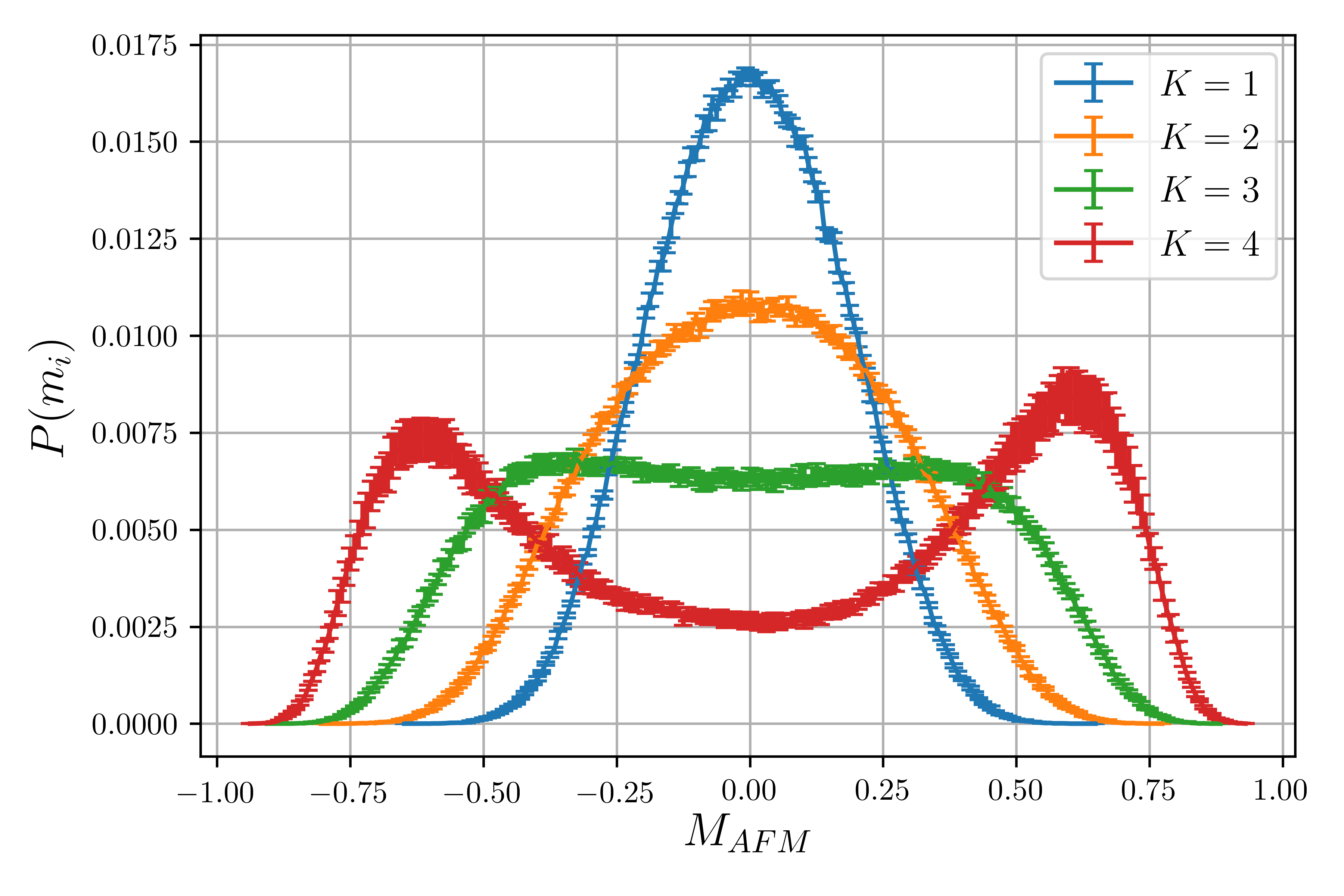}
    \includegraphics[width=0.98\columnwidth]{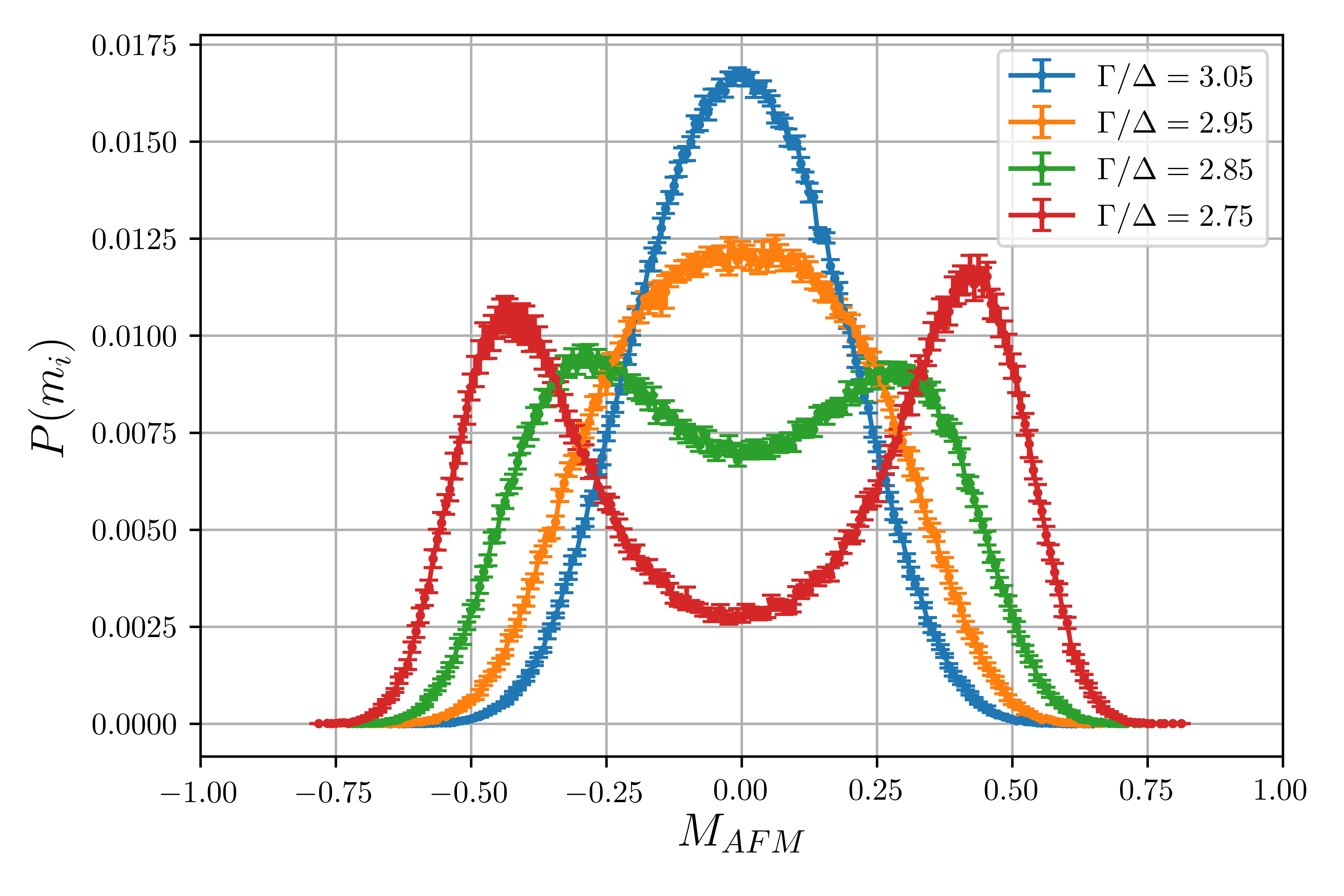}
    \caption{\textbf{Distribution of the order parameter.} (Top) Distribution at a fixed value $(\Gamma/\Delta, \Delta \beta) = (3.05, 1.645)$ with increasing embedding size. At $M_{AFM}=0$, the magnitude of the curves decrease with increasing $K$ (i.e. $K=1$ is the largest here). By increasing $K$ one moves from the paramagnetic-phase distribution centered at $m=0$ to an ordered-phase distribution with two modes at finite $\pm m$. Here, we pick \textit{every} chain (logical qubit) to be of size $K$, and in order to couple two neighbouring logical qubits (defining the native square lattice topology), we select the physical qubits randomly (out of the $K$ possible choices per chain)}. (Bottom) Distribution for $K=1$ with decreasing transverse-field ($\Delta \beta = 1.645$).
    The system size is $L=16$. At $M_{AFM}=0$, the magnitude of the curves decrease with decreasing $\Gamma/\Delta$ (i.e. $\Gamma/\Delta = 3.05$ is the largest here). Each curve is from a single MC run of at least $2^{18}$ samples. Errors computed as in Appendix \ref{sect:sim_params}.
    \label{fig:ord_param_hist}
\end{figure}

We study the order parameter $M_{AFM}$ in more detail in Fig.~\ref{fig:ord_param_hist} (top). We notice a very clear effect, that as one increases the embedding size $K$, but keeping all other parameters fixed, the system exhibits more order. In particular, the $K=1$ case is sampled just into the paramagnetic-phase, but for $K=4$ the system clearly has entered an ordered-phase.
Increasing $K$ in this setting is therefore similar to decreasing the transverse-field $\Gamma$, which is shown in Fig.~\ref{fig:ord_param_hist} (bottom), for reference. One may naturally ask therefore whether phase transitions can be obscured by such physics. To answer this one needs to study not only increasing $K$, but also increasing $N$ (ideally in the thermodynamic limit). This is the topic of the next section.

\subsection{Scaling properties}\label{sec:scaling_properties}

Here we study the effect of embedding on phase transition properties. This is relevant as it one promising use case for quantum annealers for studying physical systems  \cite{harris-tfim,king-topological, dw-shastry-sutherland, king-top-2021}. 

In the 2D anti-ferromagnet model there is a phase transition occurring at non-zero $\Gamma$, provided the temperature is low enough (n.b. the $T=0$ transition occurs at $\Gamma/\Delta = 3.04$) \cite{Elliott_1971, 2d-qmc}.

The Binder cumulant of the order parameter $g=1 - \langle M_{AFM}^4\rangle/3\langle M_{AFM}^2\rangle^2$ is a quantity that is commonly used to compute the critical point of a phase transition from numerical data in both thermal and quantum phase transitions \cite{binder:81,binder:81b,harris-tfim}. In the thermodynamic limit, $g$ is expected to be a step function of the control parameter (\ie the temperature $T$, or the transverse-field $\Gamma$): $g=2/3$ in the ordered phase and $g=0$ outside. At finite system size, the curves of the Binder cumulant will smoothly interpolate the $L\rightarrow \infty$ behaviour, but according to the finite-size-scaling (FSS) Ansatz the value of $g$ will crucially \emph{not} depend on $L$ at criticality. Thus, one way of finding the critical point is to plot the Binder cumulant curves for different system sizes and look for the point where they all intersect.

\subsubsection{Uniform embedding case}
\begin{figure}
    \centering
    \includegraphics[width=0.25\columnwidth]{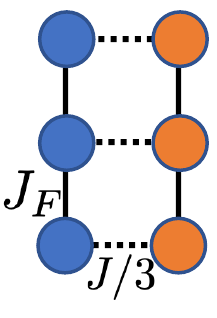}
    \caption{\textbf{Visualization of the `uniform' embedding scheme.} Here is shown the case for $K=3$, where blue (left)/orange (right) represent different logical spins. The individual inter-chain bonds (dash) are divided by $K$ in order to preserve the energy scale.}
    \label{fig:unif-embed}
\end{figure}

First, we conduct a study where the embedding chosen is completely `uniform', where in the embedded system with each chain the same length $K$ (integer), there are $K$ physical bonds between a pair of logical spins. Moreover, each spin is coupled to its `equivalent' spin at a neighbouring logical site, as shown in Fig.~\ref{fig:unif-embed}. Though this model is unpractical (as it has the connectivity to represent the native problem), it possesses very attractive properties for a numerical study. In particular, this `embedding' has a high-degree of symmetry since all logical spins/chains are essentially replicas of each other, which serves as a convenient starting point in our analysis, as we do not have worry about effects of randomness due to the embedding itself.

\begin{figure*}
    \centering
    \includegraphics[width=1.75\columnwidth]{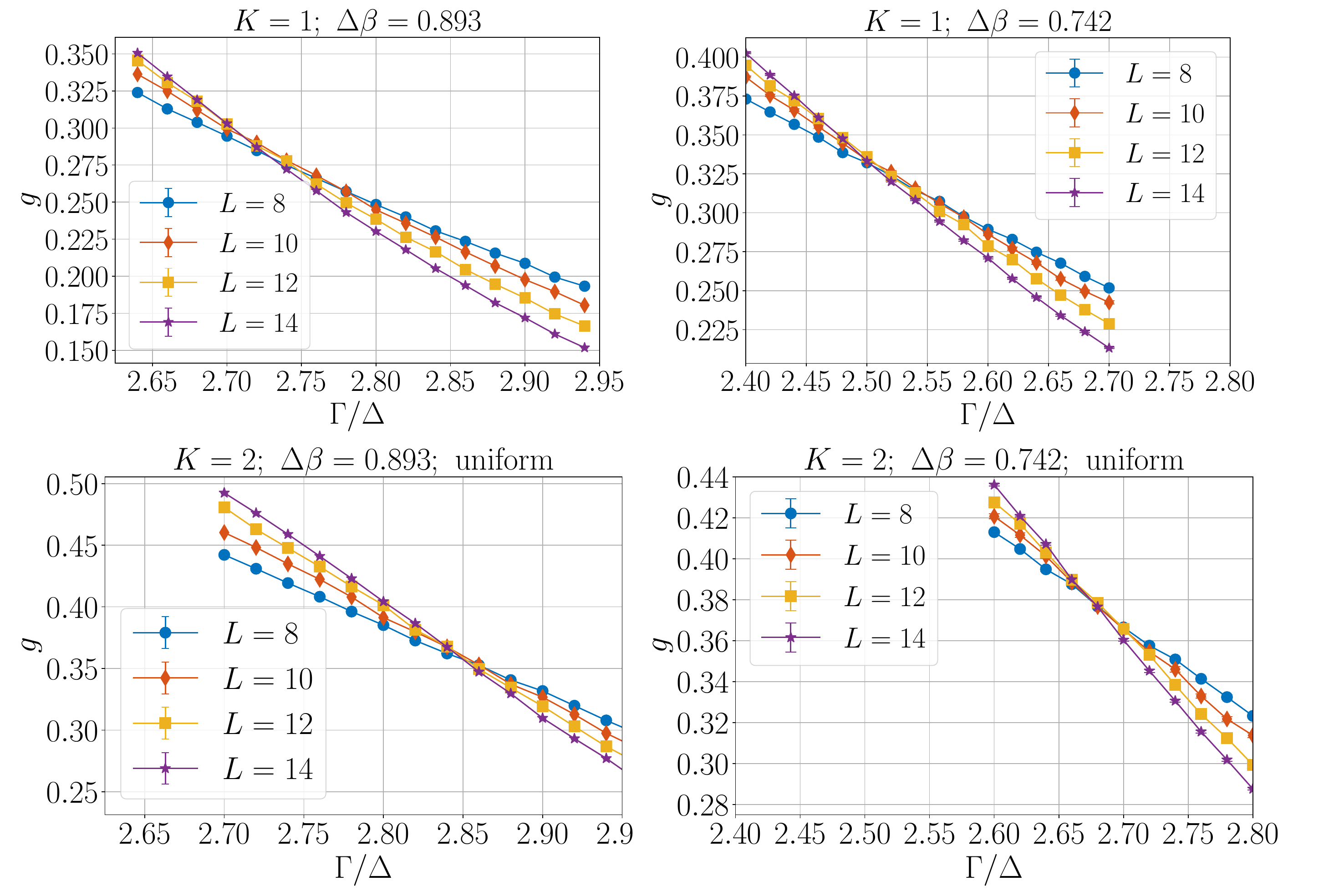}
    \caption{\textbf{Binder cumulant scaling with uniform embedding.} Here we see the dependence of the Binder cumulant $g$ on transverse field in the case of no embedding (top), and with the `uniform' embedding case for $K=2$ (bottom), for two inverse temperatures $\Delta \beta$. We see the effect of embedding changes the value of $g$ such that the location at which the curves cross is shifted in $\Gamma$. Error bars (which are smaller than the marker sizes) are the standard error of the mean over at least 20 samples (each of $2^{17}$ measurements) per point.}
    \label{fig:fss_uniform_k2}
\end{figure*}

\begin{figure*}
    \centering
    \includegraphics[width=1.75\columnwidth]{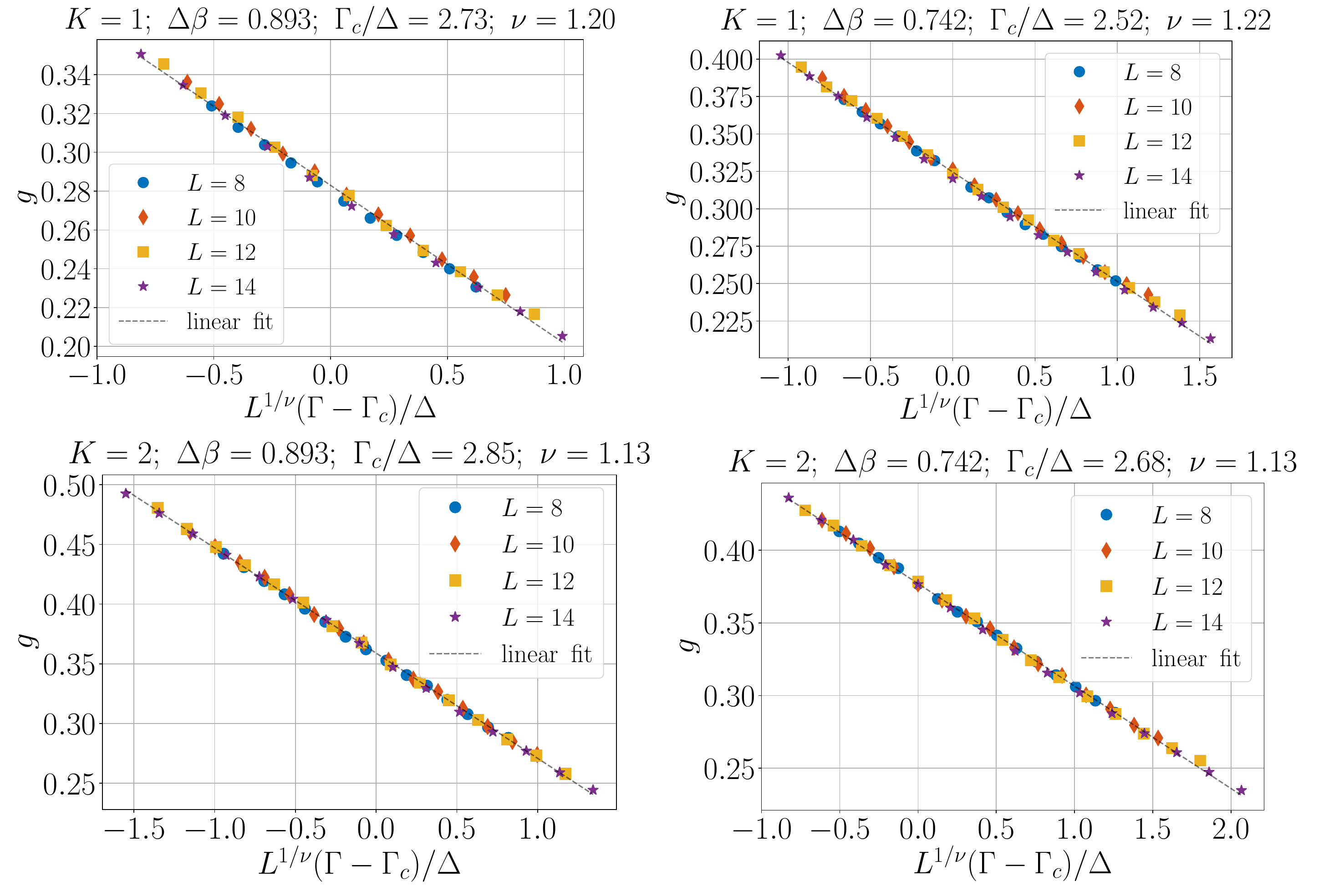}
    \caption{\textbf{FSS Ansatz of Binder cumulant data.} Data collapse of data in Fig.~\ref{fig:fss_uniform_k2}, for the `uniform' embedding. Extracted critical values $\Gamma_c, \nu$ shown in each sub-figure title, which are found by iterating over many values, and picking those which minimize the least-squares error to a linear fit.}
    \label{fig:collapse_fit}
\end{figure*}

In Figs.~\ref{fig:fss_uniform_k2}, \ref{fig:collapse_fit} we perform a FSS analysis of the Binder cumulant $g$, for an embedding of size $K=1$ (native problem) and $K=2$. The Ansatz we use is that near to the critical point $\Gamma_c$, we have $g = g(L^{1/\nu}(\Gamma - \Gamma_c))$, \ie it is scale invariant (see \eg Ref.~\cite{harris-tfim}). Interestingly, from Fig.~\ref{fig:fss_uniform_k2} we find that in this particular embedding scheme, the FSS Ansatz still appears valid (namely, there is a scale-invariant location), though the critical point is shifted to larger values of transverse-field.

\begin{figure}
    \centering
    \includegraphics[width=0.98\columnwidth]{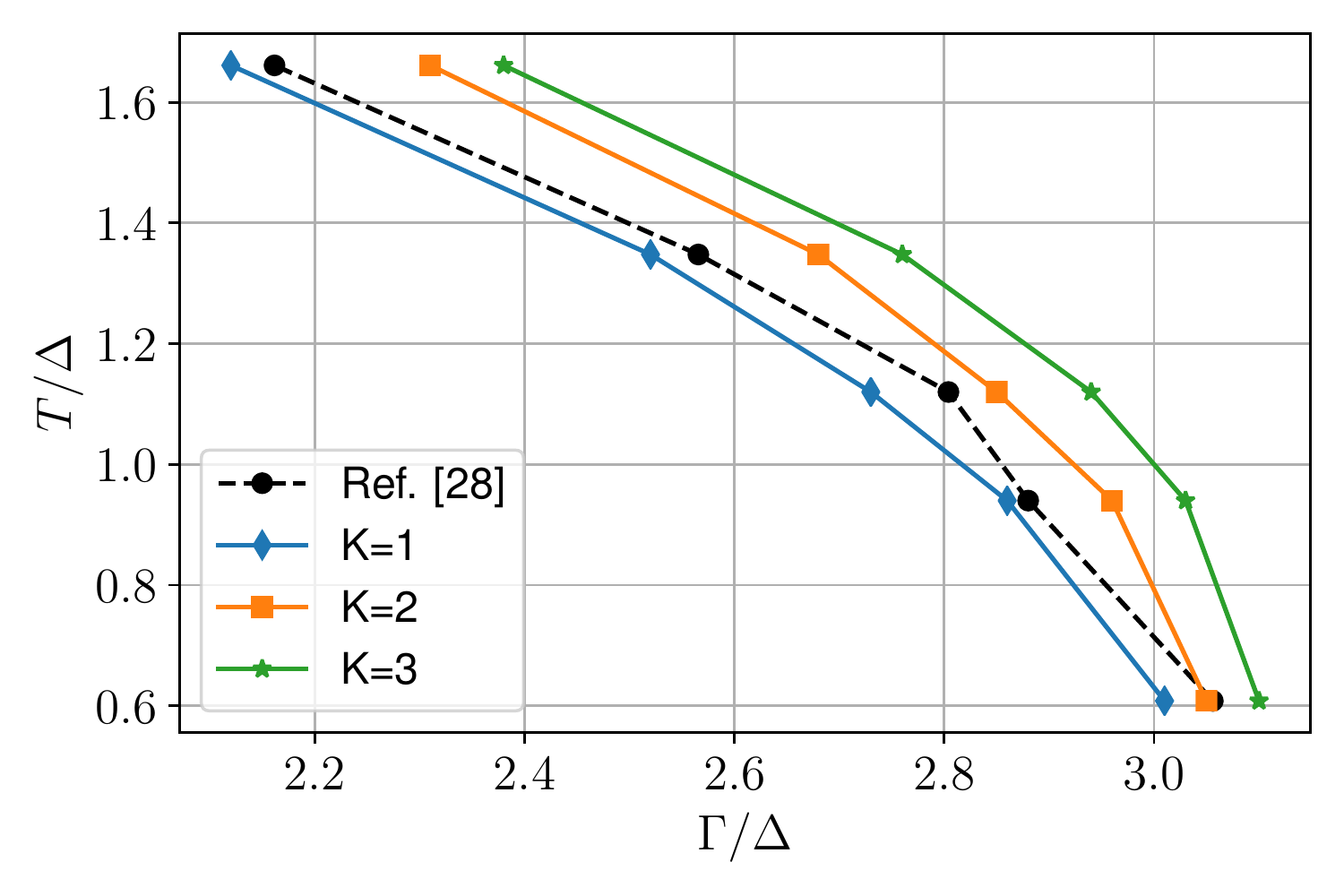}
    \caption{\textbf{Phase transition boundary shifting with increasing $K$.} Here we use the `uniform' embedding scheme, and each value in the plot is extracted by performing an analysis as in Fig.~\ref{fig:collapse_fit}. Also shown is the approximate high-temperature expansion results of Ref.~\cite{Elliott_1971} (Table 2, quadratic lattice \cite{elliott_mapping}). Note that the results of Ref.~\cite{Elliott_1971} are only approximate, and also pertain to the infinite sized system, thus explaining the small discrepancy with our numerical results (also note Ref.~\cite{Elliott_1971} overestimates the $T=0$ result from Ref.~\cite{2d-qmc}, of $\Gamma_c/\Delta = 3.04$).}
    \label{fig:phase_bdy}
\end{figure}

In Fig.~\ref{fig:collapse_fit} we collapse the data using the Ansatz above by isolating the region near to where the curves cross, and use a linear approximation to extract the critical values.
We can see visually that both for $K=1$ and $K=2$ the procedure faithfully extracts the critical value $\Gamma_c$, which can be seen by inspection of Fig.~\ref{fig:fss_uniform_k2}.

Of interest is how the phase boundary changes, which we show a portion of in Fig.~\ref{fig:phase_bdy}. We also include the case for $K=3$ here, which continues to shift the boundary to larger transverse-field. This is broadly consistent with the fully-connected $p$-spin models in quantum annealing correction (QAC) which have been studied through a mean field analysis \cite{QAC-mean-field,QAC-finite-temp, NQAC-pspin}. We will discuss this more in Sect.~\ref{sect:conclusion}.

\subsubsection{Averaging over realizations}
In the previous subsection we examined a model of embedding that exhibits a clear phase boundary in the $(T, \Gamma)$ plane, albeit with a shifted boundary from the native model. 
However, that model of embedding, whilst convenient both physically and numerically, is not a practical one for any real system, since such an embedding would have the connectivity to represent the native problem itself.
Here we consider a more physical case, where between each pair of logical spins that are coupled in the native graph, there is precisely 1 physical bond present, chosen randomly (as described in Sect.~\ref{sect:embedding}).

\begin{figure}
    \centering
    \includegraphics[width=0.98\columnwidth]{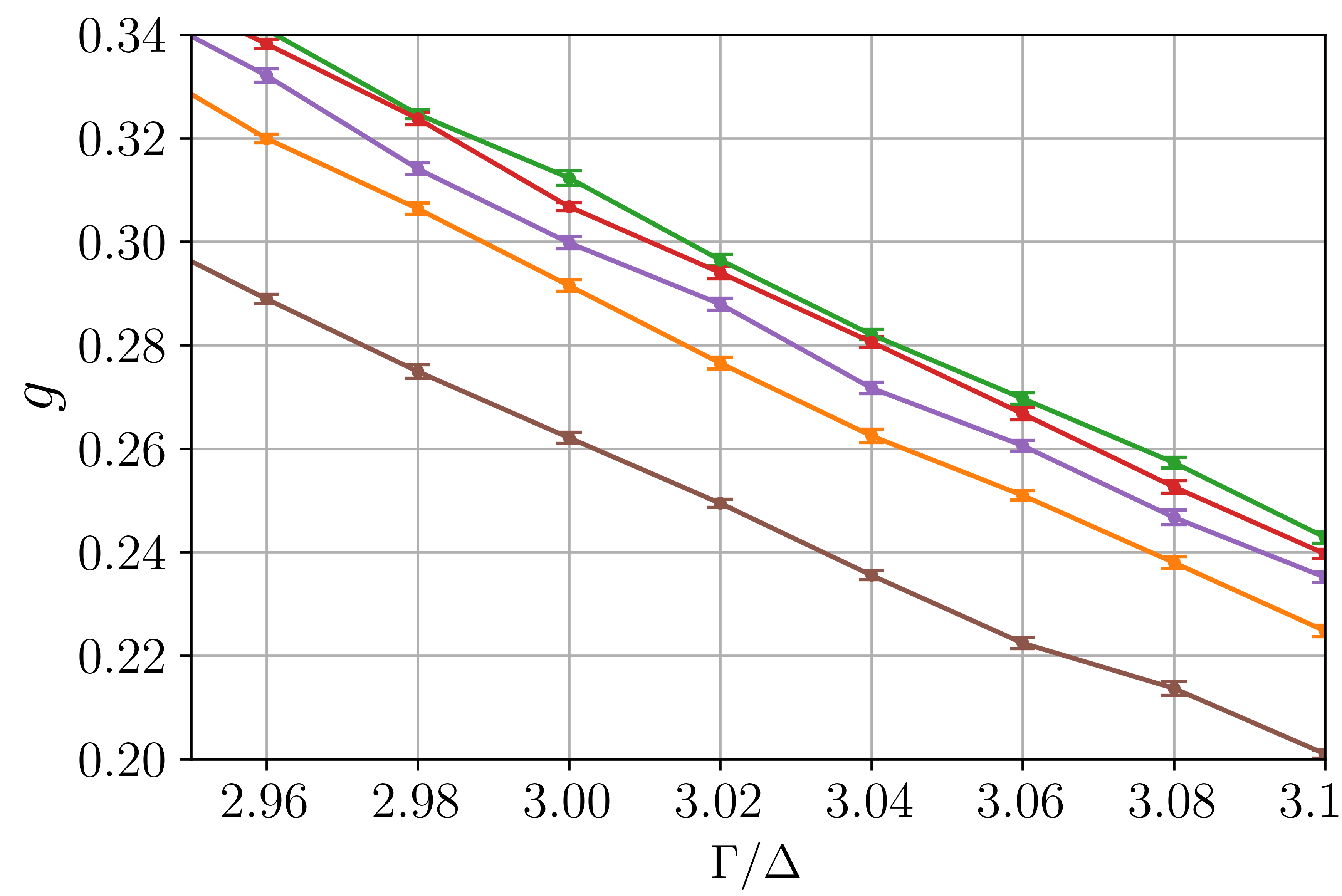}
    \caption{\textbf{Expectation value for single embedding realizations}. Here the system size is $L=10$, and each different curve corresponds to a different randomly chosen embedding, but otherwise with the same parameters ($K=2, \Delta \beta = 1.645$). We see the expectation values have large deviations between embedding realizations, indicating individual realizations are unreliable for estimating critical properties of the embedded problem class.}
    \label{fig:individual-emb}
\end{figure}

In this model we find that single realizations are unreliable for use via the FSS Ansatz, and therefore can not be used to assess the critically for this class of system. This is shown in Fig.~\ref{fig:individual-emb}, where individual embedding realizations give drastically different results.

To this end, one can consider a single embedding akin to a realization of `disorder', and to understand critical properties, one must average over sufficient realizations. Indeed, reminiscent of disorder-averaging, this approach allows us to effectively average out specific details of any single embedding, to provide a picture of the general class of system we are studying.

\begin{figure}
    \centering
    \includegraphics[width=0.98\columnwidth]{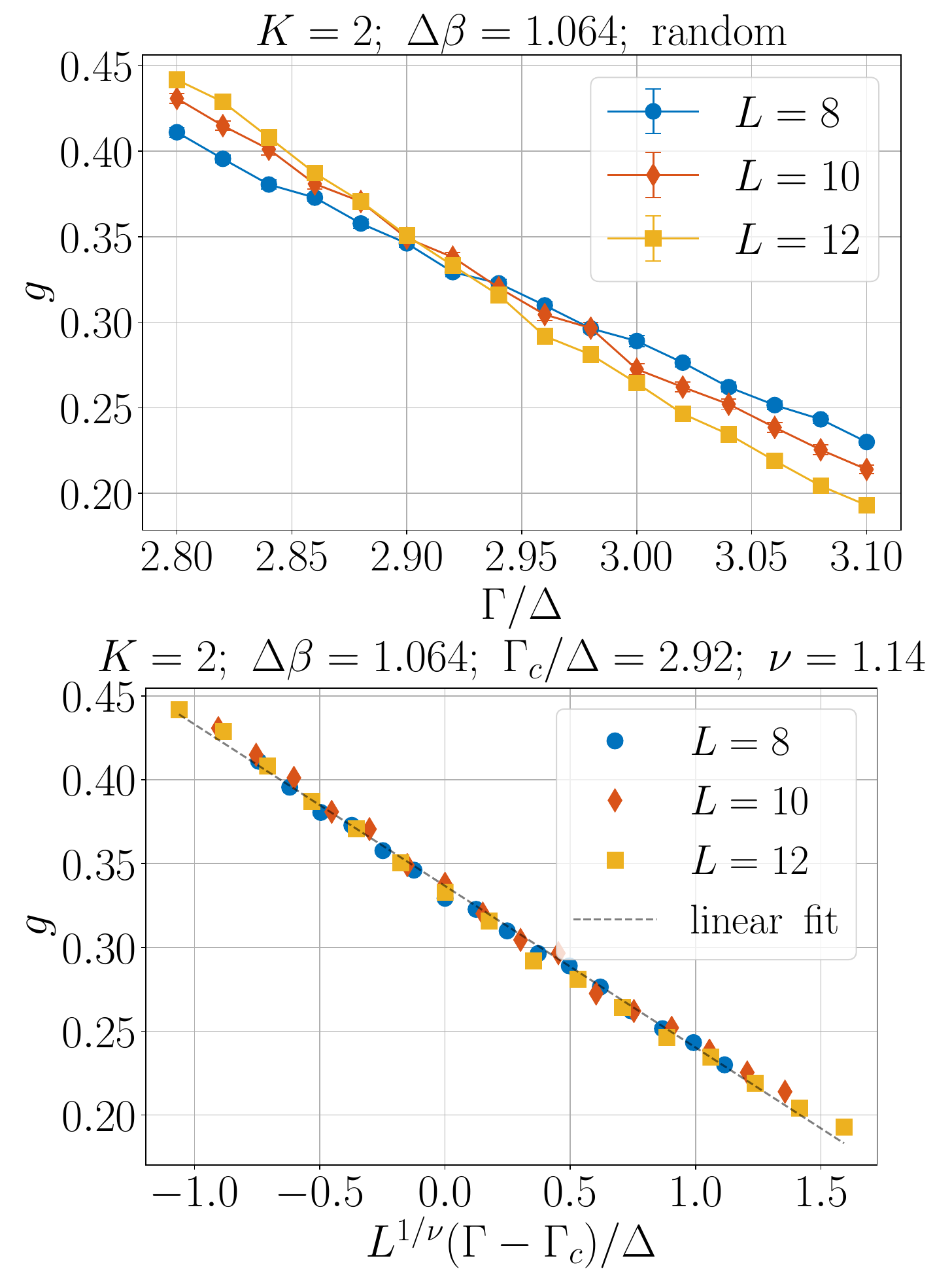}
    \caption{\textbf{Average behaviour of random embedding}. (Top) Binder cumulant scaling where points are averaged over embedding realizations.
    We can see a clear region where the curves overlap, indicating a phase transition. Note, the $K=1$ critical location is found to be $\Gamma_c/\Delta=2.86$, thus this plot indicates the embedding shifts the critical location to larger values of the transverse-field. 
    Error bars (smaller than the markers) are standard error of the mean over at least 30 independent embedding realizations per point.
    (Bottom) Data collapse of data in top figure, as described in the main text (and Fig.~\ref{fig:collapse_fit}). }
    \label{fig:embed_random_average}
\end{figure}

In this vein, we perform a similar FSS analysis as above, but on the realization-averaged Binder cumulant. In Fig.~\ref{fig:embed_random_average} we show the results for one particular temperature, where we have averaged over at least 30 embedding realizations per point. This shows that a critical region can still be seen, from which we extract an estimate for the critical point $\Gamma_c$, via the FSS Ansatz described above.

\begin{figure}
    \centering
    \includegraphics[width=0.98\columnwidth]{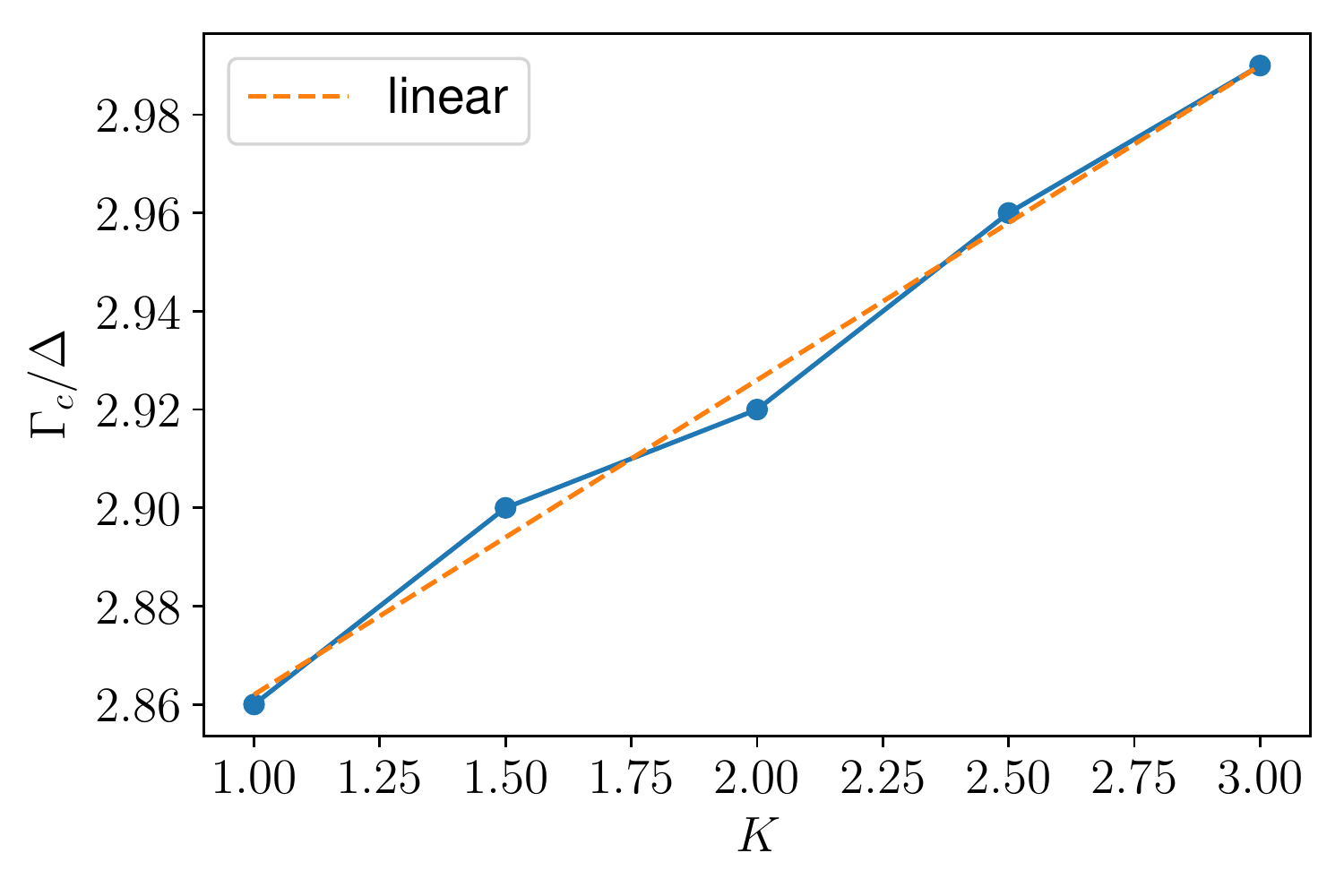}
    \caption{\textbf{Shifting of the critical location under a random embedding}. Here we perform an analysis as in Fig.~\ref{fig:embed_random_average}, for different values of embedding size $K$ (for fixed inverse temperature $\Delta \beta = 1.064$), and plot the extracted critical transverse-field value as a function of the embedding size. We fit the data to a linear fit with gradient $0.06$.}
    \label{fig:phase-trans-shift}
\end{figure}

The critical value we find here, similar to the previous subsection, is also shifted to larger values of the transverse-field. We examine this shift, taking multiple choices of the embedding size $K$ (at a fixed temperature), to see how the location of the critical point scales with the size of the embedding. In Fig.~\ref{fig:phase-trans-shift} we see that it scales linearly in $K$.

The extent to which this linear scaling holds, or for which systems is unknown, however, this observation hints at the possibility of extracting $K=1$ (\ie the native problem) critical properties, by extrapolating from larger embedding sizes, similar to how zero-noise extrapolation can be used to compute ideal (noiseless) quantum observables in the presence of noise, by artificially increasing the noise level \cite{ZNE1, ZNE2}. In the setting of our work, whilst it will generally not be possible to \textit{decrease} the embedding size all the way to $K=1$, it would often be possible to consider a range of embedding sizes $K>1$, and therefore may allow one to perform such an analysis. We leave this as a future avenue of research.

\begin{figure}
    \centering
    \includegraphics[width=0.98\columnwidth]{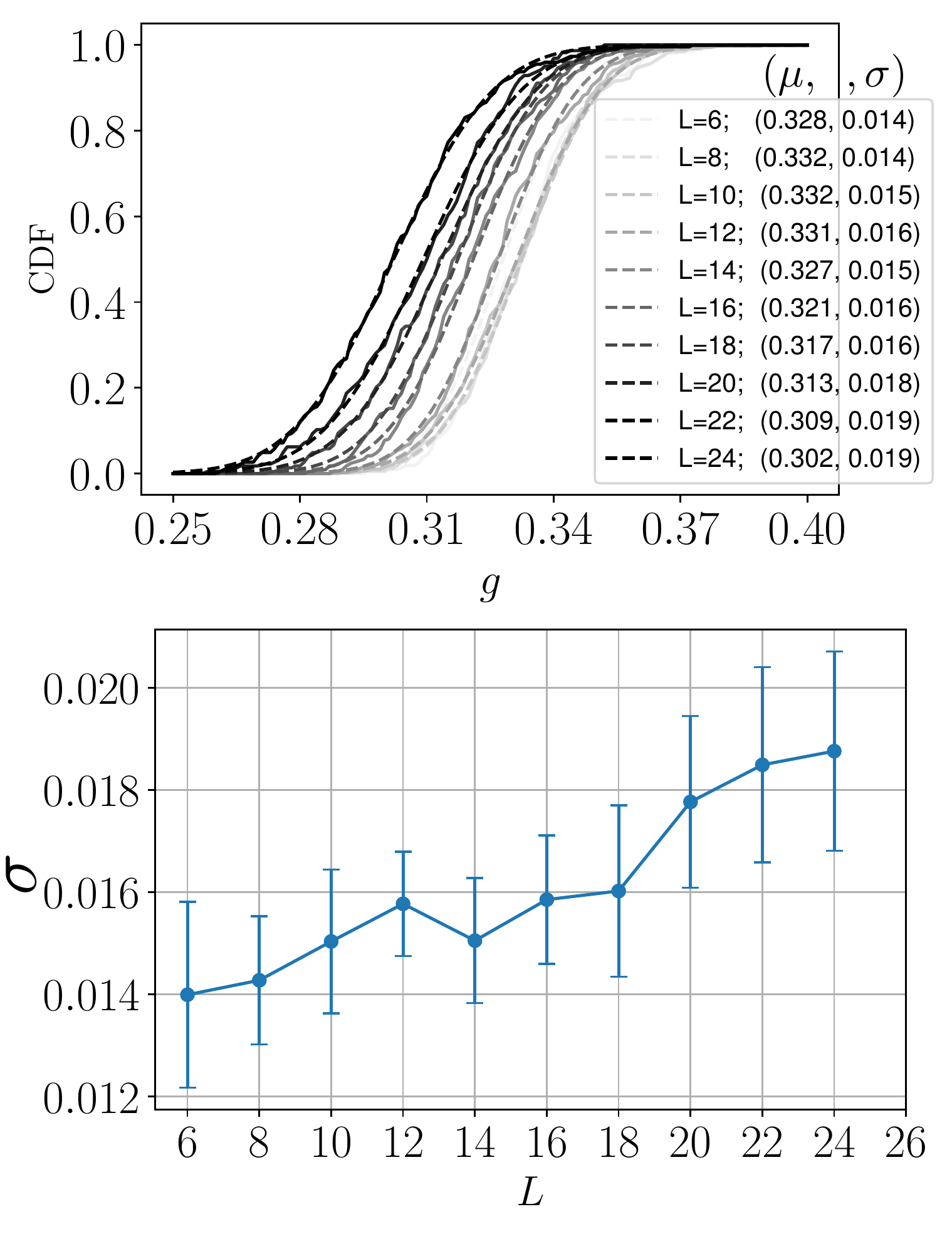}
    \caption{\textbf{Distribution  of Binder cumulant values, over embedding realizations.} Here we study a system with $K=2$, and $\Delta \beta = 1.064$, and study the distribution of the Binder cumulant, $g$, sampled near to the estimated critical point (at $\Gamma / \Delta=2.92$). (Top) Empirical Cumulative Distribution Function (CDF) of the data (solid) compared to the best-fit normal distribution CDF, with parameters in the legend (mean $\mu$, standard deviation $\sigma$). (Bottom) Extracted standard deviation $\sigma$ values, as a function of system size $L$. We take over $150$ samples (embedding realizations) per data point, and error bars represent the 95\% confidence interval by bootstrapping. Given $M$ sampled values $\{x_m\}_{m=1}^M$ one defines the empirical CDF $F(x)$ as a sum of Heaviside step functions $F(x) \equiv M^{-1}\sum_{m=1}^{M} \theta(x-x_m)$.}
    \label{fig:stdev_binder}
\end{figure}

Lastly we study how many samples are required in order to estimate a value of the Binder cumulant, as a function of problem size. In particular, since each embedding realization gives different results (see Fig.~\ref{fig:individual-emb}), the Binder cumulant over realizations gives a distribution. In Fig.~\ref{fig:stdev_binder} we compare the cumulative distribution function (CDF) of Binder cumulant data, to the CDF of a normal distribution, with parameters extracted by curve fitting. With this, we can plot the standard deviation of the distribution as a function of the (native problem) system size $L$, which shows that the standard deviation appears to increase with system size. This indicates that as larger sizes are studied, an increasing number of samples may be required.

\section{Conclusion \label{sect:conclusion}}

The aim of this work is to improve the understanding the effect of embedding on quantum thermal sampling.

To do this, we constructed a modified QMC algorithm (LC-QMC), which can much more efficiently sample the logical subspace of an embedded Hamiltonian. 
With this, we studied in detail the effects of the embedding in a physically relevant model, which can be implemented in current generation quantum hardware. 
We believe our algorithm can be very helpful in the future in understanding the effect of embedding in other relevant models.

We derived an estimate on the probability to sample the logical subspace, which extends the previous result of Ref.~\cite{perilsOfEmbedding_PRR} to the quantum regime. This will also be of practical interest when attempting to sample such systems on a physical device, as it can be used to estimate of the number of samples required.

We also observed that phase transition boundaries shift to larger values of the transverse field, where the shift increases with the size of the embedding. In the system we studied, we noticed that for embeddings with randomness (such as the distribution of the couplings), it is necessary to perform an averaging over embedding realizations, as single realizations are typically not representative of the mean. That is, we consider embedding realizations on a similar footing as disorder realizations in disordered systems. 
We found that the shift in the critical value of the transverse field scaled linearly with embedding size, indicating that it may be possible to extract native problem statistics by extrapolation (similar to the zero-noise extrapolation protocol \cite{ZNE1,ZNE2}).

Of course, there are many other considerations in practical cases with additional degrees of freedom. For example, here we considered a fixed embedding scheme, using randomly distributed chains with identical couplings $J_F$. An interesting study would be into different embedding schemes on particular hardware graphs, and tuning $J_F$ to an optimal value. 

Moreover, we did not discuss the possibility of unembedding; \ie a postprocessing scheme to map sampled states back to the logical subspace (we only considered a rejection based approach). As in Ref.~\cite{perilsOfEmbedding_PRR} we expect this can introduce additional complications, but nevertheless, it would be worthwhile to consider schemes to help remove some of the bias we found here.

Lastly we mention that although the specific case of thermal sampling considered here is perhaps most directly relevant for quantum annealing, we stress that \textit{whenever} one needs to map a system of interest to one with limited topology, similar considerations will need to be taken into account; if embedding of any kind needs to be used, questions must be asked about how faithfully the desired physics is reproduced, though the analysis will be different depending on the setting. Generally speaking, embedding related effects would be expected to have implications for any kind of analog (\ie real-time) experiment.

\acknowledgments
We are grateful for support from NASA Ames Research Center and from DARPA under IAA 8839 Annex 125. 
JM is thankful for support from NASA Academic Mission Services, Contract No. NNA16BD14C.

\newpage
~\newpage

\bibliography{refs.bib}

%merlin.mbs apsrev4-1.bst 2010-07-25 4.21a (PWD, AO, DPC) hacked
%Control: key (0)
%Control: author (0) dotless jnrlst
%Control: editor formatted (1) identically to author
%Control: production of article title (0) allowed
%Control: page (1) range
%Control: year (0) verbatim
%Control: production of eprint (0) enabled
\begin{thebibliography}{45}%
\makeatletter
\providecommand \@ifxundefined [1]{%
 \@ifx{#1\undefined}
}%
\providecommand \@ifnum [1]{%
 \ifnum #1\expandafter \@firstoftwo
 \else \expandafter \@secondoftwo
 \fi
}%
\providecommand \@ifx [1]{%
 \ifx #1\expandafter \@firstoftwo
 \else \expandafter \@secondoftwo
 \fi
}%
\providecommand \natexlab [1]{#1}%
\providecommand \enquote  [1]{``#1''}%
\providecommand \bibnamefont  [1]{#1}%
\providecommand \bibfnamefont [1]{#1}%
\providecommand \citenamefont [1]{#1}%
\providecommand \href@noop [0]{\@secondoftwo}%
\providecommand \href [0]{\begingroup \@sanitize@url \@href}%
\providecommand \@href[1]{\@@startlink{#1}\@@href}%
\providecommand \@@href[1]{\endgroup#1\@@endlink}%
\providecommand \@sanitize@url [0]{\catcode `\\12\catcode `\$12\catcode
  `\&12\catcode `\#12\catcode `\^12\catcode `\_12\catcode `\%12\relax}%
\providecommand \@@startlink[1]{}%
\providecommand \@@endlink[0]{}%
\providecommand \url  [0]{\begingroup\@sanitize@url \@url }%
\providecommand \@url [1]{\endgroup\@href {#1}{\urlprefix }}%
\providecommand \urlprefix  [0]{URL }%
\providecommand \Eprint [0]{\href }%
\providecommand \doibase [0]{http://dx.doi.org/}%
\providecommand \selectlanguage [0]{\@gobble}%
\providecommand \bibinfo  [0]{\@secondoftwo}%
\providecommand \bibfield  [0]{\@secondoftwo}%
\providecommand \translation [1]{[#1]}%
\providecommand \BibitemOpen [0]{}%
\providecommand \bibitemStop [0]{}%
\providecommand \bibitemNoStop [0]{.\EOS\space}%
\providecommand \EOS [0]{\spacefactor3000\relax}%
\providecommand \BibitemShut  [1]{\csname bibitem#1\endcsname}%
\let\auto@bib@innerbib\@empty
%</preamble>
\bibitem [{\citenamefont {Marshall}\ \emph {et~al.}(2020)\citenamefont
  {Marshall}, \citenamefont {Di~Gioacchino},\ and\ \citenamefont
  {Rieffel}}]{perilsOfEmbedding_PRR}%
  \BibitemOpen
  \bibfield  {author} {\bibinfo {author} {\bibfnamefont {J.}~\bibnamefont
  {Marshall}}, \bibinfo {author} {\bibfnamefont {A.}~\bibnamefont
  {Di~Gioacchino}}, \ and\ \bibinfo {author} {\bibfnamefont {E.~G.}\
  \bibnamefont {Rieffel}},\ }\bibfield  {title} {\enquote {\bibinfo {title}
  {Perils of embedding for sampling problems},}\ }\href {\doibase
  10.1103/PhysRevResearch.2.023020} {\bibfield  {journal} {\bibinfo  {journal}
  {Phys. Rev. Research}\ }\textbf {\bibinfo {volume} {2}},\ \bibinfo {pages}
  {023020} (\bibinfo {year} {2020})}\BibitemShut {NoStop}%
\bibitem [{\citenamefont {Apolloni}\ \emph {et~al.}(1989)\citenamefont
  {Apolloni}, \citenamefont {Carvalho},\ and\ \citenamefont
  {de~Falco}}]{Apolloni:1989qy}%
  \BibitemOpen
  \bibfield  {author} {\bibinfo {author} {\bibfnamefont {B.}~\bibnamefont
  {Apolloni}}, \bibinfo {author} {\bibfnamefont {C.}~\bibnamefont {Carvalho}},
  \ and\ \bibinfo {author} {\bibfnamefont {D.}~\bibnamefont {de~Falco}},\
  }\bibfield  {title} {\enquote {\bibinfo {title} {Quantum stochastic
  optimization},}\ }\href {\doibase
  http://dx.doi.org/10.1016/0304-4149(89)90040-9} {\bibfield  {journal}
  {\bibinfo  {journal} {Stochastic Processes and their Applications}\ }\textbf
  {\bibinfo {volume} {33}},\ \bibinfo {pages} {233} (\bibinfo {year}
  {1989})}\BibitemShut {NoStop}%
\bibitem [{\citenamefont {Kadowaki}\ and\ \citenamefont
  {Nishimori}(1998)}]{kadowaki_quantum_1998}%
  \BibitemOpen
  \bibfield  {author} {\bibinfo {author} {\bibfnamefont {T.}~\bibnamefont
  {Kadowaki}}\ and\ \bibinfo {author} {\bibfnamefont {H.}~\bibnamefont
  {Nishimori}},\ }\bibfield  {title} {\enquote {\bibinfo {title} {Quantum
  annealing in the transverse \uppercase{I}sing model},}\ }\href {\doibase
  10.1103/PhysRevE.58.5355} {\bibfield  {journal} {\bibinfo  {journal} {Phys.
  Rev. E}\ }\textbf {\bibinfo {volume} {58}},\ \bibinfo {pages} {5355}
  (\bibinfo {year} {1998})}\BibitemShut {NoStop}%
\bibitem [{\citenamefont {Smelyanskiy}\ \emph {et~al.}(2020)\citenamefont
  {Smelyanskiy}, \citenamefont {Kechedzhi}, \citenamefont {Boixo},
  \citenamefont {Isakov}, \citenamefont {Neven},\ and\ \citenamefont
  {Altshuler}}]{pop-transfer}%
  \BibitemOpen
  \bibfield  {author} {\bibinfo {author} {\bibfnamefont {V.~N.}\ \bibnamefont
  {Smelyanskiy}}, \bibinfo {author} {\bibfnamefont {K.}~\bibnamefont
  {Kechedzhi}}, \bibinfo {author} {\bibfnamefont {S.}~\bibnamefont {Boixo}},
  \bibinfo {author} {\bibfnamefont {S.~V.}\ \bibnamefont {Isakov}}, \bibinfo
  {author} {\bibfnamefont {H.}~\bibnamefont {Neven}}, \ and\ \bibinfo {author}
  {\bibfnamefont {B.}~\bibnamefont {Altshuler}},\ }\bibfield  {title} {\enquote
  {\bibinfo {title} {{Nonergodic Delocalized States for Efficient Population
  Transfer within a Narrow Band of the Energy Landscape}},}\ }\href {\doibase
  10.1103/PhysRevX.10.011017} {\bibfield  {journal} {\bibinfo  {journal} {Phys.
  Rev. X}\ }\textbf {\bibinfo {volume} {10}},\ \bibinfo {pages} {011017}
  (\bibinfo {year} {2020})}\BibitemShut {NoStop}%
\bibitem [{\citenamefont {Neill}\ \emph {et~al.}(2018)\citenamefont {Neill},
  \citenamefont {Roushan}, \citenamefont {Kechedzhi}, \citenamefont {Boixo},
  \citenamefont {Isakov}, \citenamefont {Smelyanskiy}, \citenamefont {Megrant},
  \citenamefont {Chiaro}, \citenamefont {Dunsworth}, \citenamefont {Arya},
  \citenamefont {Barends}, \citenamefont {Burkett}, \citenamefont {Chen},
  \citenamefont {Chen}, \citenamefont {Fowler}, \citenamefont {Foxen},
  \citenamefont {Giustina}, \citenamefont {Graff}, \citenamefont {Jeffrey},
  \citenamefont {Huang}, \citenamefont {Kelly}, \citenamefont {Klimov},
  \citenamefont {Lucero}, \citenamefont {Mutus}, \citenamefont {Neeley},
  \citenamefont {Quintana}, \citenamefont {Sank}, \citenamefont {Vainsencher},
  \citenamefont {Wenner}, \citenamefont {White}, \citenamefont {Neven},\ and\
  \citenamefont {Martinis}}]{blueprint-suprem}%
  \BibitemOpen
  \bibfield  {author} {\bibinfo {author} {\bibfnamefont {C.}~\bibnamefont
  {Neill}}, \bibinfo {author} {\bibfnamefont {P.}~\bibnamefont {Roushan}},
  \bibinfo {author} {\bibfnamefont {K.}~\bibnamefont {Kechedzhi}}, \bibinfo
  {author} {\bibfnamefont {S.}~\bibnamefont {Boixo}}, \bibinfo {author}
  {\bibfnamefont {S.~V.}\ \bibnamefont {Isakov}}, \bibinfo {author}
  {\bibfnamefont {V.}~\bibnamefont {Smelyanskiy}}, \bibinfo {author}
  {\bibfnamefont {A.}~\bibnamefont {Megrant}}, \bibinfo {author} {\bibfnamefont
  {B.}~\bibnamefont {Chiaro}}, \bibinfo {author} {\bibfnamefont
  {A.}~\bibnamefont {Dunsworth}}, \bibinfo {author} {\bibfnamefont
  {K.}~\bibnamefont {Arya}}, \bibinfo {author} {\bibfnamefont {R.}~\bibnamefont
  {Barends}}, \bibinfo {author} {\bibfnamefont {B.}~\bibnamefont {Burkett}},
  \bibinfo {author} {\bibfnamefont {Y.}~\bibnamefont {Chen}}, \bibinfo {author}
  {\bibfnamefont {Z.}~\bibnamefont {Chen}}, \bibinfo {author} {\bibfnamefont
  {A.}~\bibnamefont {Fowler}}, \bibinfo {author} {\bibfnamefont
  {B.}~\bibnamefont {Foxen}}, \bibinfo {author} {\bibfnamefont
  {M.}~\bibnamefont {Giustina}}, \bibinfo {author} {\bibfnamefont
  {R.}~\bibnamefont {Graff}}, \bibinfo {author} {\bibfnamefont
  {E.}~\bibnamefont {Jeffrey}}, \bibinfo {author} {\bibfnamefont
  {T.}~\bibnamefont {Huang}}, \bibinfo {author} {\bibfnamefont
  {J.}~\bibnamefont {Kelly}}, \bibinfo {author} {\bibfnamefont
  {P.}~\bibnamefont {Klimov}}, \bibinfo {author} {\bibfnamefont
  {E.}~\bibnamefont {Lucero}}, \bibinfo {author} {\bibfnamefont
  {J.}~\bibnamefont {Mutus}}, \bibinfo {author} {\bibfnamefont
  {M.}~\bibnamefont {Neeley}}, \bibinfo {author} {\bibfnamefont
  {C.}~\bibnamefont {Quintana}}, \bibinfo {author} {\bibfnamefont
  {D.}~\bibnamefont {Sank}}, \bibinfo {author} {\bibfnamefont {A.}~\bibnamefont
  {Vainsencher}}, \bibinfo {author} {\bibfnamefont {J.}~\bibnamefont {Wenner}},
  \bibinfo {author} {\bibfnamefont {T.~C.}\ \bibnamefont {White}}, \bibinfo
  {author} {\bibfnamefont {H.}~\bibnamefont {Neven}}, \ and\ \bibinfo {author}
  {\bibfnamefont {J.~M.}\ \bibnamefont {Martinis}},\ }\bibfield  {title}
  {\enquote {\bibinfo {title} {A blueprint for demonstrating quantum supremacy
  with superconducting qubits},}\ }\href {\doibase 10.1126/science.aao4309}
  {\bibfield  {journal} {\bibinfo  {journal} {Science}\ }\textbf {\bibinfo
  {volume} {360}},\ \bibinfo {pages} {195} (\bibinfo {year}
  {2018})}\BibitemShut {NoStop}%
\bibitem [{\citenamefont {Parra-Rodriguez}\ \emph {et~al.}(2020)\citenamefont
  {Parra-Rodriguez}, \citenamefont {Lougovski}, \citenamefont {Lamata},
  \citenamefont {Solano},\ and\ \citenamefont {Sanz}}]{digital-analog}%
  \BibitemOpen
  \bibfield  {author} {\bibinfo {author} {\bibfnamefont {A.}~\bibnamefont
  {Parra-Rodriguez}}, \bibinfo {author} {\bibfnamefont {P.}~\bibnamefont
  {Lougovski}}, \bibinfo {author} {\bibfnamefont {L.}~\bibnamefont {Lamata}},
  \bibinfo {author} {\bibfnamefont {E.}~\bibnamefont {Solano}}, \ and\ \bibinfo
  {author} {\bibfnamefont {M.}~\bibnamefont {Sanz}},\ }\bibfield  {title}
  {\enquote {\bibinfo {title} {Digital-analog quantum computation},}\ }\href
  {\doibase 10.1103/PhysRevA.101.022305} {\bibfield  {journal} {\bibinfo
  {journal} {Phys. Rev. A}\ }\textbf {\bibinfo {volume} {101}},\ \bibinfo
  {pages} {022305} (\bibinfo {year} {2020})}\BibitemShut {NoStop}%
\bibitem [{\citenamefont {Braum{\"u}ller}\ \emph {et~al.}(2022)\citenamefont
  {Braum{\"u}ller}, \citenamefont {Karamlou}, \citenamefont {Yanay},
  \citenamefont {Kannan}, \citenamefont {Kim}, \citenamefont {Kjaergaard},
  \citenamefont {Melville}, \citenamefont {Niedzielski}, \citenamefont {Sung},\
  and\ \citenamefont {et~al.}}]{analogue-otoc}%
  \BibitemOpen
  \bibfield  {author} {\bibinfo {author} {\bibfnamefont {J.}~\bibnamefont
  {Braum{\"u}ller}}, \bibinfo {author} {\bibfnamefont {A.~H.}\ \bibnamefont
  {Karamlou}}, \bibinfo {author} {\bibfnamefont {Y.}~\bibnamefont {Yanay}},
  \bibinfo {author} {\bibfnamefont {B.}~\bibnamefont {Kannan}}, \bibinfo
  {author} {\bibfnamefont {D.}~\bibnamefont {Kim}}, \bibinfo {author}
  {\bibfnamefont {M.}~\bibnamefont {Kjaergaard}}, \bibinfo {author}
  {\bibfnamefont {A.}~\bibnamefont {Melville}}, \bibinfo {author}
  {\bibfnamefont {B.~M.}\ \bibnamefont {Niedzielski}}, \bibinfo {author}
  {\bibfnamefont {Y.}~\bibnamefont {Sung}}, \ and\ \bibinfo {author}
  {\bibfnamefont {A.~Veps{\"a}l{\"a}inen}\ \bibnamefont {et~al.}},\ }\bibfield
  {title} {\enquote {\bibinfo {title} {{Probing quantum information propagation
  with out-of-time-ordered correlators}},}\ }\href {\doibase
  10.1038/s41567-021-01430-w} {\bibfield  {journal} {\bibinfo  {journal} {Nat.
  Phys.}\ }\textbf {\bibinfo {volume} {18}},\ \bibinfo {pages} {172} (\bibinfo
  {year} {2022})}\BibitemShut {NoStop}%
\bibitem [{\citenamefont {Choi}(2008)}]{Choi1}%
  \BibitemOpen
  \bibfield  {author} {\bibinfo {author} {\bibfnamefont {V.}~\bibnamefont
  {Choi}},\ }\bibfield  {title} {{\selectlanguage {English}\enquote {\bibinfo
  {title} {{Minor-embedding in adiabatic quantum computation: I. The parameter
  setting problem}},}\ }}\href {\doibase 10.1007/s11128-008-0082-9} {\bibfield
  {journal} {\bibinfo  {journal} {Quant. Inf. Proc.}\ }\textbf {\bibinfo
  {volume} {7}},\ \bibinfo {pages} {193} (\bibinfo {year} {2008})}\BibitemShut
  {NoStop}%
\bibitem [{\citenamefont {Ackley}\ \emph {et~al.}(1985)\citenamefont {Ackley},
  \citenamefont {Hinton},\ and\ \citenamefont {Sejnowski}}]{boltzmann-machine}%
  \BibitemOpen
  \bibfield  {author} {\bibinfo {author} {\bibfnamefont {D.~H.}\ \bibnamefont
  {Ackley}}, \bibinfo {author} {\bibfnamefont {G.~E.}\ \bibnamefont {Hinton}},
  \ and\ \bibinfo {author} {\bibfnamefont {T.~J.}\ \bibnamefont {Sejnowski}},\
  }\bibfield  {title} {\enquote {\bibinfo {title} {{A Learning Algorithm for
  Boltzmann Machines}},}\ }\href {\doibase
  https://doi.org/10.1016/S0364-0213(85)80012-4} {\bibfield  {journal}
  {\bibinfo  {journal} {Cognitive Science}\ }\textbf {\bibinfo {volume} {9}},\
  \bibinfo {pages} {147} (\bibinfo {year} {1985})}\BibitemShut {NoStop}%
\bibitem [{\citenamefont {Amin}\ \emph {et~al.}(2018)\citenamefont {Amin},
  \citenamefont {Andriyash}, \citenamefont {Rolfe}, \citenamefont
  {Kulchytskyy},\ and\ \citenamefont {Melko}}]{Amin:boltzmann}%
  \BibitemOpen
  \bibfield  {author} {\bibinfo {author} {\bibfnamefont {M.~H.}\ \bibnamefont
  {Amin}}, \bibinfo {author} {\bibfnamefont {E.}~\bibnamefont {Andriyash}},
  \bibinfo {author} {\bibfnamefont {J.}~\bibnamefont {Rolfe}}, \bibinfo
  {author} {\bibfnamefont {B.}~\bibnamefont {Kulchytskyy}}, \ and\ \bibinfo
  {author} {\bibfnamefont {R.}~\bibnamefont {Melko}},\ }\bibfield  {title}
  {\enquote {\bibinfo {title} {{Quantum Boltzmann Machine}},}\ }\href {\doibase
  10.1103/PhysRevX.8.021050} {\bibfield  {journal} {\bibinfo  {journal} {Phys.
  Rev. X}\ }\textbf {\bibinfo {volume} {8}},\ \bibinfo {pages} {021050}
  (\bibinfo {year} {2018})}\BibitemShut {NoStop}%
\bibitem [{\citenamefont {{Adachi}}\ and\ \citenamefont
  {{Henderson}}(2015)}]{adachi}%
  \BibitemOpen
  \bibfield  {author} {\bibinfo {author} {\bibfnamefont {S.~H.}\ \bibnamefont
  {{Adachi}}}\ and\ \bibinfo {author} {\bibfnamefont {M.~P.}\ \bibnamefont
  {{Henderson}}},\ }\bibfield  {title} {\enquote {\bibinfo {title}
  {{Application of Quantum Annealing to Training of Deep Neural Networks}},}\
  }\href {https://arxiv.org/abs/1510.06356} {\bibfield  {journal} {\bibinfo
  {journal} {arXiv:1510.06356}\ } (\bibinfo {year} {2015})}\BibitemShut
  {NoStop}%
\bibitem [{\citenamefont {Benedetti}\ \emph {et~al.}(2016)\citenamefont
  {Benedetti}, \citenamefont {Realpe-G\'omez}, \citenamefont {Biswas},\ and\
  \citenamefont {Perdomo-Ortiz}}]{perdomo}%
  \BibitemOpen
  \bibfield  {author} {\bibinfo {author} {\bibfnamefont {M.}~\bibnamefont
  {Benedetti}}, \bibinfo {author} {\bibfnamefont {J.}~\bibnamefont
  {Realpe-G\'omez}}, \bibinfo {author} {\bibfnamefont {R.}~\bibnamefont
  {Biswas}}, \ and\ \bibinfo {author} {\bibfnamefont {A.}~\bibnamefont
  {Perdomo-Ortiz}},\ }\bibfield  {title} {\enquote {\bibinfo {title}
  {Estimation of effective temperatures in quantum annealers for sampling
  applications: A case study with possible applications in deep learning},}\
  }\href {\doibase 10.1103/PhysRevA.94.022308} {\bibfield  {journal} {\bibinfo
  {journal} {Phys. Rev. A}\ }\textbf {\bibinfo {volume} {94}},\ \bibinfo
  {pages} {022308} (\bibinfo {year} {2016})}\BibitemShut {NoStop}%
\bibitem [{\citenamefont {Benedetti}\ \emph {et~al.}(2017)\citenamefont
  {Benedetti}, \citenamefont {Realpe-G{\'o}mez}, \citenamefont {Biswas},\ and\
  \citenamefont {Perdomo-Ortiz}}]{DW2x-gibbs-ML}%
  \BibitemOpen
  \bibfield  {author} {\bibinfo {author} {\bibfnamefont {M.}~\bibnamefont
  {Benedetti}}, \bibinfo {author} {\bibfnamefont {J.}~\bibnamefont
  {Realpe-G{\'o}mez}}, \bibinfo {author} {\bibfnamefont {R.}~\bibnamefont
  {Biswas}}, \ and\ \bibinfo {author} {\bibfnamefont {A.}~\bibnamefont
  {Perdomo-Ortiz}},\ }\bibfield  {title} {\enquote {\bibinfo {title}
  {{Quantum-Assisted Learning of Hardware-Embedded Probabilistic Graphical
  Models}},}\ }\href
  {https://journals.aps.org/prx/abstract/10.1103/PhysRevX.7.041052} {\bibfield
  {journal} {\bibinfo  {journal} {Phys. Rev. X}\ }\textbf {\bibinfo {volume}
  {7}},\ \bibinfo {pages} {041052} (\bibinfo {year} {2017})}\BibitemShut
  {NoStop}%
\bibitem [{\citenamefont {Wilson}\ \emph {et~al.}(2021)\citenamefont {Wilson},
  \citenamefont {Vandal}, \citenamefont {Hogg},\ and\ \citenamefont
  {Rieffel}}]{QAAAN}%
  \BibitemOpen
  \bibfield  {author} {\bibinfo {author} {\bibfnamefont {M.}~\bibnamefont
  {Wilson}}, \bibinfo {author} {\bibfnamefont {T.}~\bibnamefont {Vandal}},
  \bibinfo {author} {\bibfnamefont {T.}~\bibnamefont {Hogg}}, \ and\ \bibinfo
  {author} {\bibfnamefont {E.~G.}\ \bibnamefont {Rieffel}},\ }\bibfield
  {title} {\enquote {\bibinfo {title} {{Quantum-assisted associative
  adversarial network: Applying quantum annealing in deep learning}},}\ }\href
  {https://doi.org/10.1007/s42484-021-00047-9} {\bibfield  {journal} {\bibinfo
  {journal} {Quantum Machine Intelligence}\ }\textbf {\bibinfo {volume} {3}},\
  \bibinfo {pages} {19} (\bibinfo {year} {2021})}\BibitemShut {NoStop}%
\bibitem [{\citenamefont {Li}\ \emph {et~al.}(2020)\citenamefont {Li},
  \citenamefont {Albash},\ and\ \citenamefont {Lidar}}]{boltzmann-li-qac}%
  \BibitemOpen
  \bibfield  {author} {\bibinfo {author} {\bibfnamefont {R.~Y.}\ \bibnamefont
  {Li}}, \bibinfo {author} {\bibfnamefont {T.}~\bibnamefont {Albash}}, \ and\
  \bibinfo {author} {\bibfnamefont {D.~A.}\ \bibnamefont {Lidar}},\ }\bibfield
  {title} {\enquote {\bibinfo {title} {{Limitations of error corrected quantum
  annealing in improving the performance of Boltzmann machines}},}\ }\href
  {https://doi.org/10.1088/2058-9565/ab9aab} {\bibfield  {journal} {\bibinfo
  {journal} {Quantum Sci. Technol.}\ }\textbf {\bibinfo {volume} {5}},\
  \bibinfo {pages} {04501} (\bibinfo {year} {2020})}\BibitemShut {NoStop}%
\bibitem [{\citenamefont {Caldeira}\ \emph {et~al.}(2019)\citenamefont
  {Caldeira}, \citenamefont {Job}, \citenamefont {Adachi}, \citenamefont
  {Nord},\ and\ \citenamefont {Perdue}}]{boltzmann-galaxy}%
  \BibitemOpen
  \bibfield  {author} {\bibinfo {author} {\bibfnamefont {J.}~\bibnamefont
  {Caldeira}}, \bibinfo {author} {\bibfnamefont {J.}~\bibnamefont {Job}},
  \bibinfo {author} {\bibfnamefont {S.~H.}\ \bibnamefont {Adachi}}, \bibinfo
  {author} {\bibfnamefont {B.}~\bibnamefont {Nord}}, \ and\ \bibinfo {author}
  {\bibfnamefont {G.~N.}\ \bibnamefont {Perdue}},\ }\bibfield  {title}
  {\enquote {\bibinfo {title} {{Restricted Boltzmann Machines for galaxy
  morphology classification with a quantum annealer}},}\ }\href
  {https://arxiv.org/abs/1911.06259} {\bibfield  {journal} {\bibinfo  {journal}
  {arXiv:1911.06259}\ } (\bibinfo {year} {2019})}\BibitemShut {NoStop}%
\bibitem [{\citenamefont {Kairys}\ \emph {et~al.}(2020)\citenamefont {Kairys},
  \citenamefont {King}, \citenamefont {Ozfidan}, \citenamefont {Boothby},
  \citenamefont {Raymond}, \citenamefont {Banerjee},\ and\ \citenamefont
  {Humble}}]{dw-shastry-sutherland}%
  \BibitemOpen
  \bibfield  {author} {\bibinfo {author} {\bibfnamefont {P.}~\bibnamefont
  {Kairys}}, \bibinfo {author} {\bibfnamefont {A.~D.}\ \bibnamefont {King}},
  \bibinfo {author} {\bibfnamefont {I.}~\bibnamefont {Ozfidan}}, \bibinfo
  {author} {\bibfnamefont {K.}~\bibnamefont {Boothby}}, \bibinfo {author}
  {\bibfnamefont {J.}~\bibnamefont {Raymond}}, \bibinfo {author} {\bibfnamefont
  {A.}~\bibnamefont {Banerjee}}, \ and\ \bibinfo {author} {\bibfnamefont
  {T.~S.}\ \bibnamefont {Humble}},\ }\bibfield  {title} {\enquote {\bibinfo
  {title} {{Simulating the Shastry-Sutherland Ising Model Using Quantum
  Annealing}},}\ }\href {\doibase 10.1103/PRXQuantum.1.020320} {\bibfield
  {journal} {\bibinfo  {journal} {PRX Quantum}\ }\textbf {\bibinfo {volume}
  {1}},\ \bibinfo {pages} {020320} (\bibinfo {year} {2020})}\BibitemShut
  {NoStop}%
\bibitem [{\citenamefont {Khoshaman}\ \emph {et~al.}(2018)\citenamefont
  {Khoshaman}, \citenamefont {Vinci}, \citenamefont {Denis}, \citenamefont
  {Andriyash}, \citenamefont {Sadeghi},\ and\ \citenamefont {Amin}}]{qvae}%
  \BibitemOpen
  \bibfield  {author} {\bibinfo {author} {\bibfnamefont {A.}~\bibnamefont
  {Khoshaman}}, \bibinfo {author} {\bibfnamefont {W.}~\bibnamefont {Vinci}},
  \bibinfo {author} {\bibfnamefont {B.}~\bibnamefont {Denis}}, \bibinfo
  {author} {\bibfnamefont {E.}~\bibnamefont {Andriyash}}, \bibinfo {author}
  {\bibfnamefont {H.}~\bibnamefont {Sadeghi}}, \ and\ \bibinfo {author}
  {\bibfnamefont {M.~H.}\ \bibnamefont {Amin}},\ }\bibfield  {title} {\enquote
  {\bibinfo {title} {Quantum variational autoencoder},}\ }\href {\doibase
  10.1088/2058-9565/aada1f} {\bibfield  {journal} {\bibinfo  {journal} {Quant.
  Sci. and Tech.}\ }\textbf {\bibinfo {volume} {4}},\ \bibinfo {pages} {014001}
  (\bibinfo {year} {2018})}\BibitemShut {NoStop}%
\bibitem [{\citenamefont {Harris}\ \emph {et~al.}(2018)\citenamefont {Harris},
  \citenamefont {Sato}, \citenamefont {Berkley}, \citenamefont {Reis},
  \citenamefont {Altomare}, \citenamefont {Amin}, \citenamefont {Boothby},
  \citenamefont {Bunyk}, \citenamefont {Deng}, \citenamefont {Enderud},
  \citenamefont {Huang}, \citenamefont {Hoskinson}, \citenamefont {Johnson},
  \citenamefont {Ladizinsky}, \citenamefont {Ladizinsky}, \citenamefont
  {Lanting}, \citenamefont {Li}, \citenamefont {Medina}, \citenamefont
  {Molavi}, \citenamefont {Neufeld}, \citenamefont {Oh}, \citenamefont
  {Pavlov}, \citenamefont {Perminov}, \citenamefont {Poulin-Lamarre},
  \citenamefont {Rich}, \citenamefont {Smirnov}, \citenamefont {Swenson},
  \citenamefont {Tsai}, \citenamefont {Volkmann}, \citenamefont {Whittaker},\
  and\ \citenamefont {Yao}}]{harris-tfim}%
  \BibitemOpen
  \bibfield  {author} {\bibinfo {author} {\bibfnamefont {R.}~\bibnamefont
  {Harris}}, \bibinfo {author} {\bibfnamefont {Y.}~\bibnamefont {Sato}},
  \bibinfo {author} {\bibfnamefont {A.~J.}\ \bibnamefont {Berkley}}, \bibinfo
  {author} {\bibfnamefont {M.}~\bibnamefont {Reis}}, \bibinfo {author}
  {\bibfnamefont {F.}~\bibnamefont {Altomare}}, \bibinfo {author}
  {\bibfnamefont {M.~H.}\ \bibnamefont {Amin}}, \bibinfo {author}
  {\bibfnamefont {K.}~\bibnamefont {Boothby}}, \bibinfo {author} {\bibfnamefont
  {P.}~\bibnamefont {Bunyk}}, \bibinfo {author} {\bibfnamefont
  {C.}~\bibnamefont {Deng}}, \bibinfo {author} {\bibfnamefont {C.}~\bibnamefont
  {Enderud}}, \bibinfo {author} {\bibfnamefont {S.}~\bibnamefont {Huang}},
  \bibinfo {author} {\bibfnamefont {E.}~\bibnamefont {Hoskinson}}, \bibinfo
  {author} {\bibfnamefont {M.~W.}\ \bibnamefont {Johnson}}, \bibinfo {author}
  {\bibfnamefont {E.}~\bibnamefont {Ladizinsky}}, \bibinfo {author}
  {\bibfnamefont {N.}~\bibnamefont {Ladizinsky}}, \bibinfo {author}
  {\bibfnamefont {T.}~\bibnamefont {Lanting}}, \bibinfo {author} {\bibfnamefont
  {R.}~\bibnamefont {Li}}, \bibinfo {author} {\bibfnamefont {T.}~\bibnamefont
  {Medina}}, \bibinfo {author} {\bibfnamefont {R.}~\bibnamefont {Molavi}},
  \bibinfo {author} {\bibfnamefont {R.}~\bibnamefont {Neufeld}}, \bibinfo
  {author} {\bibfnamefont {T.}~\bibnamefont {Oh}}, \bibinfo {author}
  {\bibfnamefont {I.}~\bibnamefont {Pavlov}}, \bibinfo {author} {\bibfnamefont
  {I.}~\bibnamefont {Perminov}}, \bibinfo {author} {\bibfnamefont
  {G.}~\bibnamefont {Poulin-Lamarre}}, \bibinfo {author} {\bibfnamefont
  {C.}~\bibnamefont {Rich}}, \bibinfo {author} {\bibfnamefont {A.}~\bibnamefont
  {Smirnov}}, \bibinfo {author} {\bibfnamefont {L.}~\bibnamefont {Swenson}},
  \bibinfo {author} {\bibfnamefont {N.}~\bibnamefont {Tsai}}, \bibinfo {author}
  {\bibfnamefont {M.}~\bibnamefont {Volkmann}}, \bibinfo {author}
  {\bibfnamefont {J.}~\bibnamefont {Whittaker}}, \ and\ \bibinfo {author}
  {\bibfnamefont {J.}~\bibnamefont {Yao}},\ }\bibfield  {title} {\enquote
  {\bibinfo {title} {Phase transitions in a programmable quantum spin glass
  simulator},}\ }\href {\doibase 10.1126/science.aat2025} {\bibfield  {journal}
  {\bibinfo  {journal} {Science}\ }\textbf {\bibinfo {volume} {361}},\ \bibinfo
  {pages} {162} (\bibinfo {year} {2018})}\BibitemShut {NoStop}%
\bibitem [{\citenamefont {King}\ \emph {et~al.}(2018)\citenamefont {King},
  \citenamefont {Carrasquilla}, \citenamefont {Raymond}, \citenamefont
  {Ozfidan}, \citenamefont {Andriyash}, \citenamefont {Berkley}, \citenamefont
  {Reis}, \citenamefont {Lanting}, \citenamefont {Harris}, \citenamefont
  {Altomare}, \citenamefont {Boothby}, \citenamefont {Bunyk}, \citenamefont
  {Enderud}, \citenamefont {Fr{\'e}chette}, \citenamefont {Hoskinson},
  \citenamefont {Ladizinsky}, \citenamefont {Oh}, \citenamefont
  {Poulin-Lamarre}, \citenamefont {Rich}, \citenamefont {Sato}, \citenamefont
  {Smirnov}, \citenamefont {Swenson}, \citenamefont {Volkmann}, \citenamefont
  {Whittaker}, \citenamefont {Yao}, \citenamefont {Ladizinsky}, \citenamefont
  {Johnson}, \citenamefont {Hilton},\ and\ \citenamefont
  {Amin}}]{king-topological}%
  \BibitemOpen
  \bibfield  {author} {\bibinfo {author} {\bibfnamefont {A.~D.}\ \bibnamefont
  {King}}, \bibinfo {author} {\bibfnamefont {J.}~\bibnamefont {Carrasquilla}},
  \bibinfo {author} {\bibfnamefont {J.}~\bibnamefont {Raymond}}, \bibinfo
  {author} {\bibfnamefont {I.}~\bibnamefont {Ozfidan}}, \bibinfo {author}
  {\bibfnamefont {E.}~\bibnamefont {Andriyash}}, \bibinfo {author}
  {\bibfnamefont {A.}~\bibnamefont {Berkley}}, \bibinfo {author} {\bibfnamefont
  {M.}~\bibnamefont {Reis}}, \bibinfo {author} {\bibfnamefont {T.}~\bibnamefont
  {Lanting}}, \bibinfo {author} {\bibfnamefont {R.}~\bibnamefont {Harris}},
  \bibinfo {author} {\bibfnamefont {F.}~\bibnamefont {Altomare}}, \bibinfo
  {author} {\bibfnamefont {K.}~\bibnamefont {Boothby}}, \bibinfo {author}
  {\bibfnamefont {P.~I.}\ \bibnamefont {Bunyk}}, \bibinfo {author}
  {\bibfnamefont {C.}~\bibnamefont {Enderud}}, \bibinfo {author} {\bibfnamefont
  {A.}~\bibnamefont {Fr{\'e}chette}}, \bibinfo {author} {\bibfnamefont
  {E.}~\bibnamefont {Hoskinson}}, \bibinfo {author} {\bibfnamefont
  {N.}~\bibnamefont {Ladizinsky}}, \bibinfo {author} {\bibfnamefont
  {T.}~\bibnamefont {Oh}}, \bibinfo {author} {\bibfnamefont {G.}~\bibnamefont
  {Poulin-Lamarre}}, \bibinfo {author} {\bibfnamefont {C.}~\bibnamefont
  {Rich}}, \bibinfo {author} {\bibfnamefont {Y.}~\bibnamefont {Sato}}, \bibinfo
  {author} {\bibfnamefont {A.~Y.}\ \bibnamefont {Smirnov}}, \bibinfo {author}
  {\bibfnamefont {L.~J.}\ \bibnamefont {Swenson}}, \bibinfo {author}
  {\bibfnamefont {M.~H.}\ \bibnamefont {Volkmann}}, \bibinfo {author}
  {\bibfnamefont {J.}~\bibnamefont {Whittaker}}, \bibinfo {author}
  {\bibfnamefont {J.}~\bibnamefont {Yao}}, \bibinfo {author} {\bibfnamefont
  {E.}~\bibnamefont {Ladizinsky}}, \bibinfo {author} {\bibfnamefont {M.~W.}\
  \bibnamefont {Johnson}}, \bibinfo {author} {\bibfnamefont {J.}~\bibnamefont
  {Hilton}}, \ and\ \bibinfo {author} {\bibfnamefont {M.~H.}\ \bibnamefont
  {Amin}},\ }\bibfield  {title} {\enquote {\bibinfo {title} {Observation of
  topological phenomena in a programmable lattice of 1,800 qubits},}\ }\href
  {https://doi.org/10.1038/s41586-018-0410-x} {\bibfield  {journal} {\bibinfo
  {journal} {Nature}\ }\textbf {\bibinfo {volume} {560}},\ \bibinfo {pages}
  {456} (\bibinfo {year} {2018})}\BibitemShut {NoStop}%
\bibitem [{\citenamefont {Gonzalez~Izquierdo}\ \emph
  {et~al.}(2021)\citenamefont {Gonzalez~Izquierdo}, \citenamefont {Hen},\ and\
  \citenamefont {Albash}}]{izquierdo2020testing}%
  \BibitemOpen
  \bibfield  {author} {\bibinfo {author} {\bibfnamefont {Z.}~\bibnamefont
  {Gonzalez~Izquierdo}}, \bibinfo {author} {\bibfnamefont {I.}~\bibnamefont
  {Hen}}, \ and\ \bibinfo {author} {\bibfnamefont {T.}~\bibnamefont {Albash}},\
  }\bibfield  {title} {\enquote {\bibinfo {title} {{Testing a Quantum Annealer
  as a Quantum Thermal Sampler}},}\ }\href {\doibase 10.1145/3464456}
  {\bibfield  {journal} {\bibinfo  {journal} {ACM Transactions on Quantum
  Computing}\ }\textbf {\bibinfo {volume} {2}} (\bibinfo {year} {2021}),\
  10.1145/3464456}\BibitemShut {NoStop}%
\bibitem [{\citenamefont {Nishimura}\ \emph {et~al.}(2020)\citenamefont
  {Nishimura}, \citenamefont {Nishimori},\ and\ \citenamefont
  {Katzgraber}}]{Griffiths-McCoy}%
  \BibitemOpen
  \bibfield  {author} {\bibinfo {author} {\bibfnamefont {K.}~\bibnamefont
  {Nishimura}}, \bibinfo {author} {\bibfnamefont {H.}~\bibnamefont
  {Nishimori}}, \ and\ \bibinfo {author} {\bibfnamefont {H.~G.}\ \bibnamefont
  {Katzgraber}},\ }\bibfield  {title} {\enquote {\bibinfo {title}
  {{Griffiths-McCoy singularity on the diluted Chimera graph: Monte Carlo
  simulations and experiments on quantum hardware}},}\ }\href {\doibase
  10.1103/PhysRevA.102.042403} {\bibfield  {journal} {\bibinfo  {journal}
  {Phys. Rev. A}\ }\textbf {\bibinfo {volume} {102}},\ \bibinfo {pages}
  {042403} (\bibinfo {year} {2020})}\BibitemShut {NoStop}%
\bibitem [{\citenamefont {King}\ \emph {et~al.}(2021)\citenamefont {King},
  \citenamefont {Raymond}, \citenamefont {Lanting}, \citenamefont {Isakov},
  \citenamefont {Mohseni}, \citenamefont {Poulin-Lamarre}, \citenamefont
  {Ejtemaee}, \citenamefont {Bernoudy}, \citenamefont {Ozfidan}, \citenamefont
  {Smirnov},\ and\ \citenamefont {et~al.}}]{king-top-2021}%
  \BibitemOpen
  \bibfield  {author} {\bibinfo {author} {\bibfnamefont {A.~D.}\ \bibnamefont
  {King}}, \bibinfo {author} {\bibfnamefont {J.}~\bibnamefont {Raymond}},
  \bibinfo {author} {\bibfnamefont {T.}~\bibnamefont {Lanting}}, \bibinfo
  {author} {\bibfnamefont {S.~V.}\ \bibnamefont {Isakov}}, \bibinfo {author}
  {\bibfnamefont {M.}~\bibnamefont {Mohseni}}, \bibinfo {author} {\bibfnamefont
  {G.}~\bibnamefont {Poulin-Lamarre}}, \bibinfo {author} {\bibfnamefont
  {S.}~\bibnamefont {Ejtemaee}}, \bibinfo {author} {\bibfnamefont
  {W.}~\bibnamefont {Bernoudy}}, \bibinfo {author} {\bibfnamefont
  {I.}~\bibnamefont {Ozfidan}}, \bibinfo {author} {\bibfnamefont {A.~Yu.}\
  \bibnamefont {Smirnov}}, \ and\ \bibinfo {author} {\bibfnamefont {M.~Reis}\
  \bibnamefont {et~al.}},\ }\bibfield  {title} {\enquote {\bibinfo {title}
  {{Scaling advantage over path-integral Monte Carlo in quantum simulation of
  geometrically frustrated magnets}},}\ }\href {\doibase
  10.1038/s41467-021-20901-5} {\bibfield  {journal} {\bibinfo  {journal} {Nat.
  Comm.}\ }\textbf {\bibinfo {volume} {12}},\ \bibinfo {pages} {1113} (\bibinfo
  {year} {2021})}\BibitemShut {NoStop}%
\bibitem [{\citenamefont {Hastings}(2013)}]{hastings-obstructions}%
  \BibitemOpen
  \bibfield  {author} {\bibinfo {author} {\bibfnamefont {M.~B.}\ \bibnamefont
  {Hastings}},\ }\bibfield  {title} {\enquote {\bibinfo {title} {Obstructions
  to classically simulating the quantum adiabatic algorithm},}\ }\href
  {https://dl.acm.org/doi/10.5555/2535639.2535647} {\bibfield  {journal}
  {\bibinfo  {journal} {Quantum Info. Comput.}\ }\textbf {\bibinfo {volume}
  {13}},\ \bibinfo {pages} {1038} (\bibinfo {year} {2013})}\BibitemShut
  {NoStop}%
\bibitem [{\citenamefont {Raymond}\ \emph {et~al.}(2020)\citenamefont
  {Raymond}, \citenamefont {Ndiaye}, \citenamefont {Rayaprolu},\ and\
  \citenamefont {King}}]{logical-qubit-dwave}%
  \BibitemOpen
  \bibfield  {author} {\bibinfo {author} {\bibfnamefont {J.}~\bibnamefont
  {Raymond}}, \bibinfo {author} {\bibfnamefont {N.}~\bibnamefont {Ndiaye}},
  \bibinfo {author} {\bibfnamefont {G.}~\bibnamefont {Rayaprolu}}, \ and\
  \bibinfo {author} {\bibfnamefont {A.~D.}\ \bibnamefont {King}},\ }\bibfield
  {title} {\enquote {\bibinfo {title} {Improving performance of logical qubits
  by parameter tuning and topology compensation},}\ }in\ \href {\doibase
  10.1109/QCE49297.2020.00044} {\emph {\bibinfo {booktitle} {2020 IEEE
  International Conference on Quantum Computing and Engineering (QCE)}}}\
  (\bibinfo {year} {2020})\ pp.\ \bibinfo {pages} {295--305}\BibitemShut
  {NoStop}%
\bibitem [{\citenamefont {Boothby}\ \emph {et~al.}(2016)\citenamefont
  {Boothby}, \citenamefont {King},\ and\ \citenamefont {Roy}}]{dw-fast-clique}%
  \BibitemOpen
  \bibfield  {author} {\bibinfo {author} {\bibfnamefont {K.}~\bibnamefont
  {Boothby}}, \bibinfo {author} {\bibfnamefont {A.~D.}\ \bibnamefont {King}}, \
  and\ \bibinfo {author} {\bibfnamefont {A.}~\bibnamefont {Roy}},\ }\bibfield
  {title} {\enquote {\bibinfo {title} {{Fast clique minor generation in Chimera
  qubit connectivity graphs}},}\ }\href
  {https://doi.org/10.1007/s11128-015-1150-6} {\bibfield  {journal} {\bibinfo
  {journal} {Quantum Information Processing}\ }\textbf {\bibinfo {volume}
  {15}},\ \bibinfo {pages} {495} (\bibinfo {year} {2016})}\BibitemShut
  {NoStop}%
\bibitem [{\citenamefont {Reiner}\ \emph {et~al.}(2016)\citenamefont {Reiner},
  \citenamefont {Marthaler}, \citenamefont {Braum\"uller}, \citenamefont
  {Weides},\ and\ \citenamefont {Sch\"on}}]{fermi-hubbard-1d}%
  \BibitemOpen
  \bibfield  {author} {\bibinfo {author} {\bibfnamefont {J.-M.}\ \bibnamefont
  {Reiner}}, \bibinfo {author} {\bibfnamefont {M.}~\bibnamefont {Marthaler}},
  \bibinfo {author} {\bibfnamefont {J.}~\bibnamefont {Braum\"uller}}, \bibinfo
  {author} {\bibfnamefont {M.}~\bibnamefont {Weides}}, \ and\ \bibinfo {author}
  {\bibfnamefont {G.}~\bibnamefont {Sch\"on}},\ }\bibfield  {title} {\enquote
  {\bibinfo {title} {{Emulating the one-dimensional Fermi-Hubbard model by a
  double chain of qubits}},}\ }\href {\doibase 10.1103/PhysRevA.94.032338}
  {\bibfield  {journal} {\bibinfo  {journal} {Phys. Rev. A}\ }\textbf {\bibinfo
  {volume} {94}},\ \bibinfo {pages} {032338} (\bibinfo {year}
  {2016})}\BibitemShut {NoStop}%
\bibitem [{\citenamefont {Elliott}\ and\ \citenamefont
  {Wood}(1971)}]{Elliott_1971}%
  \BibitemOpen
  \bibfield  {author} {\bibinfo {author} {\bibfnamefont {R.~J.}\ \bibnamefont
  {Elliott}}\ and\ \bibinfo {author} {\bibfnamefont {C.}~\bibnamefont {Wood}},\
  }\bibfield  {title} {\enquote {\bibinfo {title} {{The Ising model with a
  transverse field. I. High temperature expansion}},}\ }\href {\doibase
  10.1088/0022-3719/4/15/023} {\bibfield  {journal} {\bibinfo  {journal} {J. of
  Phys. C}\ }\textbf {\bibinfo {volume} {4}},\ \bibinfo {pages} {2359}
  (\bibinfo {year} {1971})}\BibitemShut {NoStop}%
\bibitem [{\citenamefont {Rieger}\ and\ \citenamefont
  {Kawashima}(1999)}]{2d-qmc}%
  \BibitemOpen
  \bibfield  {author} {\bibinfo {author} {\bibfnamefont {H.}~\bibnamefont
  {Rieger}}\ and\ \bibinfo {author} {\bibfnamefont {N.}~\bibnamefont
  {Kawashima}},\ }\bibfield  {title} {\enquote {\bibinfo {title} {{Application
  of a continuous time cluster algorithm to the two-dimensional random quantum
  Ising ferromagnet}},}\ }\href {https://doi.org/10.1007/s100510050761}
  {\bibfield  {journal} {\bibinfo  {journal} {Eur. Phys. J. B}\ }\textbf
  {\bibinfo {volume} {9}},\ \bibinfo {pages} {233} (\bibinfo {year}
  {1999})}\BibitemShut {NoStop}%
\bibitem [{\citenamefont {Andriyash}\ and\ \citenamefont
  {Amin}(2017)}]{qmc_qa}%
  \BibitemOpen
  \bibfield  {author} {\bibinfo {author} {\bibfnamefont {E.}~\bibnamefont
  {Andriyash}}\ and\ \bibinfo {author} {\bibfnamefont {M.~H.}\ \bibnamefont
  {Amin}},\ }\bibfield  {title} {\enquote {\bibinfo {title} {{Can quantum Monte
  Carlo simulate quantum annealing?}}}\ }\href
  {https://arxiv.org/abs/1703.09277} {\bibfield  {journal} {\bibinfo  {journal}
  {arXiv:1703.09277}\ } (\bibinfo {year} {2017})}\BibitemShut {NoStop}%
\bibitem [{\citenamefont {Boixo}\ \emph {et~al.}(2014)\citenamefont {Boixo},
  \citenamefont {R{\o}nnow}, \citenamefont {Isakov}, \citenamefont {Wang},
  \citenamefont {Wecker}, \citenamefont {Lidar}, \citenamefont {Martinis},\
  and\ \citenamefont {Troyer}}]{boixo:14}%
  \BibitemOpen
  \bibfield  {author} {\bibinfo {author} {\bibfnamefont {S.}~\bibnamefont
  {Boixo}}, \bibinfo {author} {\bibfnamefont {T.~F.}\ \bibnamefont
  {R{\o}nnow}}, \bibinfo {author} {\bibfnamefont {S.~V.}\ \bibnamefont
  {Isakov}}, \bibinfo {author} {\bibfnamefont {Z.}~\bibnamefont {Wang}},
  \bibinfo {author} {\bibfnamefont {D.}~\bibnamefont {Wecker}}, \bibinfo
  {author} {\bibfnamefont {D.~A.}\ \bibnamefont {Lidar}}, \bibinfo {author}
  {\bibfnamefont {J.~M.}\ \bibnamefont {Martinis}}, \ and\ \bibinfo {author}
  {\bibfnamefont {M.}~\bibnamefont {Troyer}},\ }\bibfield  {title} {\enquote
  {\bibinfo {title} {Evidence for quantum annealing with more than one hundred
  qubits},}\ }\href {http://dx.doi.org/10.1038/nphys2900} {\bibfield  {journal}
  {\bibinfo  {journal} {Nat. Phys.}\ }\textbf {\bibinfo {volume} {10}},\
  \bibinfo {pages} {218} (\bibinfo {year} {2014})}\BibitemShut {NoStop}%
\bibitem [{\citenamefont {Marto\ifmmode~\check{n}\else \v{n}\fi{}\'ak}\ \emph
  {et~al.}(2002)\citenamefont {Marto\ifmmode~\check{n}\else \v{n}\fi{}\'ak},
  \citenamefont {Santoro},\ and\ \citenamefont {Tosatti}}]{SQA-original}%
  \BibitemOpen
  \bibfield  {author} {\bibinfo {author} {\bibfnamefont {R.}~\bibnamefont
  {Marto\ifmmode~\check{n}\else \v{n}\fi{}\'ak}}, \bibinfo {author}
  {\bibfnamefont {G.~E.}\ \bibnamefont {Santoro}}, \ and\ \bibinfo {author}
  {\bibfnamefont {E.}~\bibnamefont {Tosatti}},\ }\bibfield  {title} {\enquote
  {\bibinfo {title} {{Quantum annealing by the path-integral Monte Carlo
  method: The two-dimensional random Ising model}},}\ }\href {\doibase
  10.1103/PhysRevB.66.094203} {\bibfield  {journal} {\bibinfo  {journal} {Phys.
  Rev. B}\ }\textbf {\bibinfo {volume} {66}},\ \bibinfo {pages} {094203}
  (\bibinfo {year} {2002})}\BibitemShut {NoStop}%
\bibitem [{\citenamefont {{Crosson}}\ and\ \citenamefont
  {{Harrow}}(2016)}]{Crosson-SQA}%
  \BibitemOpen
  \bibfield  {author} {\bibinfo {author} {\bibfnamefont {E.}~\bibnamefont
  {{Crosson}}}\ and\ \bibinfo {author} {\bibfnamefont {A.~W.}\ \bibnamefont
  {{Harrow}}},\ }\bibfield  {title} {\enquote {\bibinfo {title} {{Simulated
  Quantum Annealing Can Be Exponentially Faster Than Classical Simulated
  Annealing}},}\ }in\ \href {\doibase 10.1109/FOCS.2016.81} {\emph {\bibinfo
  {booktitle} {2016 IEEE 57th Annual Symposium on Foundations of Computer
  Science (FOCS)}}}\ (\bibinfo {year} {2016})\ pp.\ \bibinfo {pages}
  {714--723}\BibitemShut {NoStop}%
\bibitem [{tem()}]{temp-note}%
  \BibitemOpen
  \href@noop {} {}\bibinfo {note} {{This temperature corresponds to the
  smallest temperature considered in Ref.~\cite{Elliott_1971} (Table 2 for the
  quadratic lattice). Also see \cite{elliott_mapping}.}}\BibitemShut {Stop}%
\bibitem [{\citenamefont {Binder}(1981{\natexlab{a}})}]{binder:81}%
  \BibitemOpen
  \bibfield  {author} {\bibinfo {author} {\bibfnamefont {K.}~\bibnamefont
  {Binder}},\ }\bibfield  {title} {\enquote {\bibinfo {title} {{Critical
  Properties from Monte Carlo Coarse Graining and Renormalization}},}\ }\href
  {\doibase 10.1103/PhysRevLett.47.693} {\bibfield  {journal} {\bibinfo
  {journal} {Phys. Rev. Lett.}\ }\textbf {\bibinfo {volume} {47}},\ \bibinfo
  {pages} {693} (\bibinfo {year} {1981}{\natexlab{a}})}\BibitemShut {NoStop}%
\bibitem [{\citenamefont {Binder}(1981{\natexlab{b}})}]{binder:81b}%
  \BibitemOpen
  \bibfield  {author} {\bibinfo {author} {\bibfnamefont {K.}~\bibnamefont
  {Binder}},\ }\bibfield  {title} {\enquote {\bibinfo {title} {Finite size
  scaling analysis of {I}sing model block distribution functions},}\ }\href
  {\doibase https://doi.org/10.1007/BF01293604} {\bibfield  {journal} {\bibinfo
   {journal} {Z. Phys. B}\ }\textbf {\bibinfo {volume} {43}},\ \bibinfo {pages}
  {119} (\bibinfo {year} {1981}{\natexlab{b}})}\BibitemShut {NoStop}%
\bibitem [{ell()}]{elliott_mapping}%
  \BibitemOpen
  \href@noop {} {}\bibinfo {note} {{Since Ref.~\cite{Elliott_1971} uses a spin
  $1/2$ convention, there is a factor of 2 difference in the definition of
  $\Gamma$ and factor of 4 difference in the definition of $T$, as compared to
  the notation in the present work.}}\BibitemShut {Stop}%
\bibitem [{\citenamefont {Matsuura}\ \emph {et~al.}(2016)\citenamefont
  {Matsuura}, \citenamefont {Nishimori}, \citenamefont {Albash},\ and\
  \citenamefont {Lidar}}]{QAC-mean-field}%
  \BibitemOpen
  \bibfield  {author} {\bibinfo {author} {\bibfnamefont {S.}~\bibnamefont
  {Matsuura}}, \bibinfo {author} {\bibfnamefont {H.}~\bibnamefont {Nishimori}},
  \bibinfo {author} {\bibfnamefont {T.}~\bibnamefont {Albash}}, \ and\ \bibinfo
  {author} {\bibfnamefont {D.~A.}\ \bibnamefont {Lidar}},\ }\bibfield  {title}
  {\enquote {\bibinfo {title} {Mean field analysis of quantum annealing
  correction},}\ }\href {\doibase 10.1103/PhysRevLett.116.220501} {\bibfield
  {journal} {\bibinfo  {journal} {Phys. Rev. Lett.}\ }\textbf {\bibinfo
  {volume} {116}},\ \bibinfo {pages} {220501} (\bibinfo {year}
  {2016})}\BibitemShut {NoStop}%
\bibitem [{\citenamefont {Matsuura}\ \emph {et~al.}(2017)\citenamefont
  {Matsuura}, \citenamefont {Nishimori}, \citenamefont {Vinci}, \citenamefont
  {Albash},\ and\ \citenamefont {Lidar}}]{QAC-finite-temp}%
  \BibitemOpen
  \bibfield  {author} {\bibinfo {author} {\bibfnamefont {S.}~\bibnamefont
  {Matsuura}}, \bibinfo {author} {\bibfnamefont {H.}~\bibnamefont {Nishimori}},
  \bibinfo {author} {\bibfnamefont {W.}~\bibnamefont {Vinci}}, \bibinfo
  {author} {\bibfnamefont {T.}~\bibnamefont {Albash}}, \ and\ \bibinfo {author}
  {\bibfnamefont {D.~A.}\ \bibnamefont {Lidar}},\ }\bibfield  {title} {\enquote
  {\bibinfo {title} {{Quantum-annealing correction at finite temperature:
  Ferromagnetic $p$-spin models}},}\ }\href {\doibase
  10.1103/PhysRevA.95.022308} {\bibfield  {journal} {\bibinfo  {journal} {Phys.
  Rev. A}\ }\textbf {\bibinfo {volume} {95}},\ \bibinfo {pages} {022308}
  (\bibinfo {year} {2017})}\BibitemShut {NoStop}%
\bibitem [{\citenamefont {Matsuura}\ \emph {et~al.}(2019)\citenamefont
  {Matsuura}, \citenamefont {Nishimori}, \citenamefont {Vinci},\ and\
  \citenamefont {Lidar}}]{NQAC-pspin}%
  \BibitemOpen
  \bibfield  {author} {\bibinfo {author} {\bibfnamefont {S.}~\bibnamefont
  {Matsuura}}, \bibinfo {author} {\bibfnamefont {H.}~\bibnamefont {Nishimori}},
  \bibinfo {author} {\bibfnamefont {W.}~\bibnamefont {Vinci}}, \ and\ \bibinfo
  {author} {\bibfnamefont {D.~A.}\ \bibnamefont {Lidar}},\ }\bibfield  {title}
  {\enquote {\bibinfo {title} {Nested quantum annealing correction at finite
  temperature: $p$-spin models},}\ }\href {\doibase 10.1103/PhysRevA.99.062307}
  {\bibfield  {journal} {\bibinfo  {journal} {Phys. Rev. A}\ }\textbf {\bibinfo
  {volume} {99}},\ \bibinfo {pages} {062307} (\bibinfo {year}
  {2019})}\BibitemShut {NoStop}%
\bibitem [{\citenamefont {Temme}\ \emph {et~al.}(2017)\citenamefont {Temme},
  \citenamefont {Bravyi},\ and\ \citenamefont {Gambetta}}]{ZNE1}%
  \BibitemOpen
  \bibfield  {author} {\bibinfo {author} {\bibfnamefont {K.}~\bibnamefont
  {Temme}}, \bibinfo {author} {\bibfnamefont {S.}~\bibnamefont {Bravyi}}, \
  and\ \bibinfo {author} {\bibfnamefont {J.~M.}\ \bibnamefont {Gambetta}},\
  }\bibfield  {title} {\enquote {\bibinfo {title} {{Error Mitigation for
  Short-Depth Quantum Circuits}},}\ }\href {\doibase
  10.1103/PhysRevLett.119.180509} {\bibfield  {journal} {\bibinfo  {journal}
  {Phys. Rev. Lett.}\ }\textbf {\bibinfo {volume} {119}},\ \bibinfo {pages}
  {180509} (\bibinfo {year} {2017})}\BibitemShut {NoStop}%
\bibitem [{\citenamefont {Li}\ and\ \citenamefont {Benjamin}(2017)}]{ZNE2}%
  \BibitemOpen
  \bibfield  {author} {\bibinfo {author} {\bibfnamefont {Y.}~\bibnamefont
  {Li}}\ and\ \bibinfo {author} {\bibfnamefont {S.~C.}\ \bibnamefont
  {Benjamin}},\ }\bibfield  {title} {\enquote {\bibinfo {title} {{Efficient
  Variational Quantum Simulator Incorporating Active Error Minimization}},}\
  }\href {\doibase 10.1103/PhysRevX.7.021050} {\bibfield  {journal} {\bibinfo
  {journal} {Phys. Rev. X}\ }\textbf {\bibinfo {volume} {7}},\ \bibinfo {pages}
  {021050} (\bibinfo {year} {2017})}\BibitemShut {NoStop}%
\bibitem [{Note1()}]{Note1}%
  \BibitemOpen
  \bibinfo {note} {In the context of the main text, $n = NK$, where $N$ is the
  native system size, and $K$ the embedding size, as defined in Sect.~\ref
  {sect:embedding}}\BibitemShut {NoStop}%
\bibitem [{Note2()}]{Note2}%
  \BibitemOpen
  \bibinfo {note} {{\protect \it i.e. }by Trotterizing the imaginary-time
  propagation of $\beta $ into $\ell $ equal steps of $\beta _{eff} \equiv
  \beta /\ell $, for some fixed positive integer $\ell $. See {\protect \it
  e.g. }\cite {SQA-original} for details.}\BibitemShut {Stop}%
\bibitem [{\citenamefont {Ambegaokar}\ and\ \citenamefont
  {Troyer}(2010)}]{mc-errors}%
  \BibitemOpen
  \bibfield  {author} {\bibinfo {author} {\bibfnamefont {V.}~\bibnamefont
  {Ambegaokar}}\ and\ \bibinfo {author} {\bibfnamefont {M.}~\bibnamefont
  {Troyer}},\ }\bibfield  {title} {\enquote {\bibinfo {title} {{Estimating
  errors reliably in Monte Carlo simulations of the Ehrenfest model}},}\ }\href
  {https://doi.org/10.1119/1.3247985} {\bibfield  {journal} {\bibinfo
  {journal} {Am. J. Phys.}\ }\textbf {\bibinfo {volume} {78}},\ \bibinfo
  {pages} {150} (\bibinfo {year} {2010})}\BibitemShut {NoStop}%
\end{thebibliography}%

\appendix

\section{Logically constrained QMC \label{sect:appendix-qmc}}

In this Appendix we describe the details of the Quantum Monte Carlo protocol we use to sample directly in the logical subspace. Let $H = H_P - \Gamma \sum_i \sigma_i^x$ be a transverse-field Hamiltonian over $n$ quantum $1/2$-spins. The $H_P$ term is assumed to be diagonal in the $\sigma^z$ product basis $\ket{\config}$, and we take it to represent the Hamiltonian of a classical combinatorial problem which has been embedded into some fixed hardware topology \footnote{In the context of the main text, $n = NK$, where $N$ is the native system size, and $K$ the embedding size, as defined in Sect.~\ref{sect:embedding}}. Therefore, $H_P$ includes the auxiliary spins and couplings coming from the embedding. We indicate with $\{\sigma_i\}$ a configuration of $n$ classical spins $\sigma_i \in \{\pm 1\}$ for $i = 1,\ldots,n$, and with $\Omega$ the set of all such configurations. Let $M$ be an observable that is diagonal in this basis
$$
M = \sum_{\config \in \Omega} m\Big(\config\Big)\;\ket{\config}\bra{\config}.
$$
The thermal average of $M$ at inverse temperature $\beta$ is given by
\begin{eqnarray*}
\avg{M} &\equiv& \frac{1}{Z}\tr\Big(M e^{-\beta H}\Big)\\ &=& \frac{1}{Z} \sum_{\config \in \Omega} m\Big(\config\Big)\; \iiprod{\config}{e^{-\beta H}}{\config} \\
&\equiv& \sum_{\config \in \Omega} m\Big(\config\Big)\; p\Big(\config\Big)
\end{eqnarray*}
where  $Z =\sum_{\omega \in \Omega}\langle \omega | e^{-\beta H }|\omega\rangle$. We have used the fact that since $H$ is stoquastic, its Boltzmann weights $\iiprod{\config}{e^{-\beta H}}{\config}$ can be normalized to a \emph{bona fide} probability distribution $p(\config)$ over $\Omega$. Given a subset of the computational basis vectors $\Omega_0 \subseteq \Omega$ (which up to this point is completely arbitrary but we will later choose to be  the set of logical configurations of the classical embedded system), the conditional probability induced by $p$ over $\Omega_0$ is
$$
p_0\Big(\config\Big) = \begin{cases} \frac{1}{Z_0} p\Big(\config\Big) & \text{if $\config \in \Omega_0$}\\
0 & \text{otherwise.}
\end{cases}
$$
where $Z_0 = \sum_{\config \in \Omega_0} p\Big(\config\Big)$ is a normalization constant (that is, $Z_0=\sum_{\omega \in \Omega_0}\langle \omega | e^{-\beta H }|\omega\rangle$). The goal will be to devise a Quantum Monte Carlo algorithm to efficiently sample $p_0$.

We can write the average value of $M$ with respect to the probability distribution $p_0$ as a sum over $\Omega_0$
\begin{eqnarray*}
\avg{M}_0 &\equiv& \sum_{\config \in \Omega_0} m\Big(\config\Big)\;p_0\Big(\config\Big) \\
&=& \frac{1}{ZZ_0}\sum_{\config \in \Omega_0} m\Big(\config\Big)\;\iiprod{\config}{e^{-\beta H}}{\config}.
\end{eqnarray*}
Now we can expand the braket $\iiprod{\config}{e^{-\beta H}}{\config}$ factor using the standard path-integral approach \footnote{\ie by Trotterizing the imaginary-time propagation of $\beta$ into $\ell$ equal steps of $\beta_{eff} \equiv \beta/\ell$, for some fixed positive integer $\ell$. See \eg \cite{SQA-original} for details.}
$$
\iiprod{\config}{e^{-\beta H}}{\config} \approx C^{n\ell} \sum_{\tau=1}^{\ell-1} \sum_{\confign{\tau} \in \Omega} e^{-\beta_{\eff} H_{\eff}\Big(\config, \confign{\tau}_{\tau}\Big)}
$$
to write the braket term a sum of Boltzmann weights with $\ell-1$ intermediate timeslices and a classical effective Hamiltonian $H_{\eff}$ over $n\ell$ classical spins. Here $ \beta_{\eff} \equiv \beta/\ell$ is an effective (inverse) temperature, $C \equiv \sqrt{\frac{1}{2} \sinh(2\beta_{\eff}\Gamma)}$ is a constant and $H_{\eff}$ has the form
$$
H_{\eff} = \sum_{\tau=0}^{\ell-1} H_P(\confign{\tau}) - J^\perp \sum_{\tau=0}^{\ell-1} \sum_{i=1}^n \sigma_i^{(\tau)} \sigma_i^{(\tau+1)}
$$
where $J^{\perp} = -\frac{1}{2\beta_{eff}}\log \tanh \beta_{\eff}\Gamma > 0$ defines a ferromagnetic interaction. Note there are periodic boundary conditions in the imaginary direction ($\sigma_i^{(\ell)} = \sigma_i^{(0)}$). If we plug this in the Equation above (and rename $\config \rightarrow \confign{0}$) then we get
\begin{widetext}
\begin{eqnarray*}
\avg{M}_0 &\approx& \frac{C^{n\ell}}{ZZ_0}\sum_{\confign{0} \in \Omega_0} \sum_{\confign{1} \in \Omega} \cdots \sum_{\confign{\ell-1} \in \Omega} m\Big(\confign{0}\Big)\;\exp\Big[-\beta_{\eff} H_{\eff}\Big(\confign{0}, \confign{1}, \ldots, \confign{\ell-1}\Big) \Big]\\
&\equiv& \sum_{\confign{0} \in \Omega_0} \sum_{\confign{1} \in \Omega} \cdots \sum_{\confign{\ell-1} \in \Omega} m\Big(\confign{0}\Big)\; P_0\Big(\confign{0}, \confign{1}, \ldots, \confign{\ell-1}\Big).
\end{eqnarray*}
\end{widetext}
Note that
\begin{enumerate}
	\item the sum over the configurations of the $0$-th timeslice only involves configurations in the restricted subset $\Omega_0$, while the sums of all the other timeslices are over the full configuration space $\Omega$.
	\item The diagonal matrix elements $\iiprod{\config}{M}{\config} = m(\config)$ of the observable $M$ that appear in the sum depend only on the configurations of the $0$-th timeslice.
\end{enumerate}
This average value $\avg{M}_0$ can be computed via Monte Carlo if we define a Markov Chain that converges in distribution to $p_0$ in the infinite-time limit. We present here a minor modification of the Wolff cluster update (with a Metropolis-Hastings acceptance probability) that achieves this. 

\subsection*{Cluster Update in Imaginary Time}
Here we prove that an appropriately designed imaginary-time only Wolff cluster update -- defined on a path-integral extended lattice of $\mathcal{N} \equiv n\ell$ classical spins -- satisfies detailed balance, whilst preserving the 0-th timeslice state to always be in $\Omega_0$ (which, as mentioned above, we will take to be the set of logical configurations). The effective Hamiltonian $H_{\eff}$ defines a percolation model on the classical spin system by connecting two spins $\sigma_i^{(\tau)},\sigma_j^{(\tau')}$ through a ``bond'' if and only if they are coupled by $H_{\eff}$. At each Monte Carlo move one creates a percolation cluster by probabilistically declaring these bonds to be active or inactive according to the following prescription. Fix a probability $p_\add \in [0,1]$ that we will explain later how to choose appropriately. Starting with a \emph{logical} spin configuration $\mu = \{\sigma_i^{(\tau)} \mid i,\tau\}\in \Omega_0 \times \Omega^{\ell-1}$ (that is, a configuration where there are no broken chains in the $\tau=0$ timeslice), the cluster move proceeds as follows:
\begin{enumerate}
	\item choose one spin $\sigma_i^{(\tau)}$ at random out of the $\mathcal{N}$ spins in the system
	\item start growing the cluster around $\sigma_i^{(\tau)}$ by defining an incident bond \emph{in the imaginary time direction} ($\langle \tau,\tau'\rangle$ for $\tau' = \tau \pm 1$) to be active with probability
	$$
		p = \begin{cases} p_\add & \text{if $\sigma_i^{(\tau)} = \sigma_i^{(\tau')}$} \\ 0 & \text{otherwise} \end{cases},
	$$
	and a bond $\langle i,j \rangle$ in the real-space direction between adjacent spins $\sigma_i^{(\tau)}$ and $\sigma_j^{(\tau)}$ to be active with probability
	$$
	p = \begin{cases}
	1 & \text{if $\tau=0$ and $\sigma_i^z,\sigma_j^z$ are physical spins}\\ & \text{representing the same logical qubit} \\
	0 & \text{otherwise}
	\end{cases}
	$$
	In case the bond is active, include the neighbouring spin $\sigma_{i'}^{(\tau')}$ in the cluster.
	\item continue growing the cluster in both the real-space and the imaginary-time directions until you close the cluster on all sides by hitting inactive bonds.
\end{enumerate}

This creates clusters in the imaginary time direction, with spatial connections where a cluster straddles the $\tau=0$ slice through the chains from an embedding. An example of step 2 is shown in Fig.~\ref{fig:lc-qmc}.

Now flip all the spins in the cluster (changing the configuration $\mu$ to the new configuration $\nu$) with probability
\begin{equation}\label{eq:metropolis}
p = \min \Big( 1, e^{-\beta_{\eff}(\tilde{E}_\nu-\tilde{E}_\mu)} \Big)
\end{equation}
where $\tilde{E}_\mu$ is the energy due \emph{only to the real-space interactions}, of state $\mu$ (and analogously $\tilde{E}_\nu$).

We will study two configurations $\mu,\nu \in \Omega_0 \times \Omega^{\ell-1}$ connected by one of these cluster spin flips. Now, a bit of notation
\begin{itemize}
	\item $C$ is the cluster, $N_c$ is the number of imaginary-time bonds connecting two spins which are both inside of the cluster.
	\item $\partial C$ is the number of \emph{ imaginary-time bonds} across the boundary of the cluster.
	\item $\partial_\mu$ is the number of bonds $\langle i,j \rangle \in \partial C$ such that $\sigma_i = \sigma_j$ in the configuration $\mu$. Analogously for $\partial_\nu$ and the configuration $\nu$
\end{itemize}
Now, since the Hamiltonian is
$$
H_\eff = H_{\mathrm{realspace}} - J^{\perp} \sum_{\tau}\sum_{i} \sigma_i^{(\tau)}\sigma_i^{(\tau+1)}
$$
then the energy of the configuration $\mu$ is
\begin{eqnarray*}
E_\mu &=& \tilde{E}_\mu - J^{\perp}N_c - 2J^{\perp}\partial_\mu + J^{\perp}\partial C \\
&&+ \textsl{other Trotter interactions}
\end{eqnarray*}
where $\tilde{E}_\mu$ is the energy due to the real-space interactions. Analogously for $\nu$
\begin{eqnarray*}
E_\nu &=& \tilde{E}_\nu - J^{\perp}N_c - 2J^{\perp}\partial_\nu + J^{\perp}\partial C \\
&&+ \textsl{other Trotter interactions}.
\end{eqnarray*}
Then for the probability distribution 
$$
\pi(\mu) \equiv \mathcal{Z}^{-1}\exp(- \beta_{\eff} H_\eff(\mu))
$$
we have that
\begin{equation}\label{eq:deltaenergy}
\frac{\pi(\mu)}{\pi(\nu)} = e^{-\beta_{\eff}(E_\mu-E_\nu)} = e^{- \beta_{\eff} (\tilde{E}_\mu - \tilde{E}_\nu)} e^{2\beta_{\eff} J^{\perp}(\partial_\mu - \partial_\nu)}
\end{equation}
Now we compute the probability of proposing the configuration $\nu$ if we start from the configuration $\mu$, \ie the probability of generating the cluster $C$.
\begin{eqnarray*}
	&&g(\mu \rightarrow \nu) = \mathrm{Pr}\Big[\;\text{choosing one of the spins in $C$ at step 1.}\;\Big] \\
	&&\quad \times \; \mathrm{Pr}\Big[\;\text{declaring all the internal bonds of $C$ as active}\;\Big]\\
	&&\quad \times \; \mathrm{Pr}\Big[\;\text{declaring the boundary bonds of $C$ as inactive}\;\Big]
\end{eqnarray*}
which gives
$$
g(\mu \rightarrow \nu) = \frac{|C|}{\mathcal{N}}(p_\add)^{N_c}(1-p_\add)^{\partial \mu}
$$
and
$$
g(\nu \rightarrow \mu) = \frac{|C|}{\mathcal{N}}(p_\add)^{N_c}(1-p_\add)^{\partial \nu}
$$
so that their ratio is
$$
\frac{g(\mu \rightarrow \nu)}{g(\nu \rightarrow \mu)} = (1-p_\add)^{\partial \mu -\partial \nu}.
$$
If we choose $p_\add \equiv 1 - \exp(-2\beta_{eff} J^{\perp})$ then we have 
\begin{equation}\label{eq:fixpadd}
\frac{g(\mu \rightarrow \nu)}{g(\nu \rightarrow \mu)} = e^{-2\beta_{\eff} J^{\perp}(\partial \mu -\partial \nu)}.
\end{equation}

We introduce some notation for better readability
$$
W_{\mu\nu} \equiv \frac{\pi(\mu)}{\pi(\nu)}, \quad\quad G_{\mu\nu} \equiv \frac{g(\mu \rightarrow \nu)}{g(\nu \rightarrow \mu)}
$$
and
$$
\tilde{W}_{\mu\nu} \equiv e^{- \beta_{\eff} (\tilde{E}_\mu - \tilde{E}_\nu)}
$$
so we have that
$$
W_{\mu\nu} = W_{\nu\mu}^{-1}, \quad\quad \tilde{W}_{\mu\nu} = \tilde{W}_{\nu\mu}^{-1}
$$
and
$$
G_{\mu\nu} = G_{\nu\mu}^{-1}, \quad\quad W_{\mu\nu} = \tilde{W}_{\mu\nu}G_{\nu\mu}
$$
See Eqs. \eqref{eq:deltaenergy} and \eqref{eq:fixpadd} for the last property. Now, detailed balance requires that 
\begin{eqnarray}\label{eq:detailedbalance}
1 &=& \frac{\pi(\mu)}{\pi(\nu)} \frac{P(\mu \rightarrow \nu)}{P(\nu \rightarrow \mu)} = \frac{\pi(\mu)}{\pi(\nu)} \frac{g(\mu \rightarrow \nu)}{g(\nu \rightarrow \mu)} \frac{A(\mu \rightarrow \nu)}{A(\nu \rightarrow \mu)} \\
&=& W_{\mu\nu} G_{\mu\nu} \frac{A(\mu \rightarrow \nu)}{A(\nu \rightarrow \mu)} = \tilde{W}_{\mu\nu} \frac{A(\mu \rightarrow \nu)}{A(\nu \rightarrow \mu)}
\end{eqnarray}
where $A(\mu \rightarrow \nu)$ is the acceptance probability of the proposed move $\mu \rightarrow \nu$. Metropolis-Hastings gives
$$
A(\mu \rightarrow \nu) \equiv \min\Big(1, \frac{\pi(\nu)g(\nu \rightarrow \mu)}{\pi(\mu)g(\mu \rightarrow \nu)} \Big) = \min\Big(1, \tilde{W}_{\nu\mu}\Big)
$$
(note: this is exactly Eq. \eqref{eq:metropolis}) so Eq. \eqref{eq:detailedbalance} becomes
$$
1 = \tilde{W}_{\mu\nu} \frac{\min\Big(1, \tilde{W}_{\nu\mu}\Big)}{\min\Big(1, \tilde{W}_{\mu\nu}\Big)}.
$$
If $\tilde{W}_{\nu\mu} =1$ this is trivially true. If $\tilde{W}_{\nu\mu} < 1$ then $\tilde{W}_{\mu\nu} = 1/\tilde{W}_{\nu\mu} > 1$ and we have
$$
\tilde{W}_{\mu\nu} \frac{\tilde{W}_{\nu\mu}}{1} = 1.
$$
If $\tilde{W}_{\nu\mu} > 1$ then $\tilde{W}_{\mu\nu} < 1$ and we get
$$
\tilde{W}_{\mu\nu}\frac{1}{\tilde{W}_{\mu\nu}} = 1
$$
so detailed balance is satisfied. If the initial state is chosen to be logical, \ie no chains are broken in the $\tau=0$ timeslice, then the Markov Chain is ergodic over the set $\Omega_0 \times \Omega^{\ell-1}$ of \emph{logical} spin configurations of the effective spin system.

The implementation of the the above described \textit{Logically-Constrained Quantum Monte-Carlo} (LC-QMC) algorithm that we used in the main text creates all clusters in the imaginary-time direction first, and then joins across the $\tau=0$ slice before attempting to flip them sequentially (so it is actually closer to a Swendsen-Wang recipe), but is otherwise identical to the process given above.

\begin{figure}
    \centering
    \includegraphics[width=0.98\columnwidth]{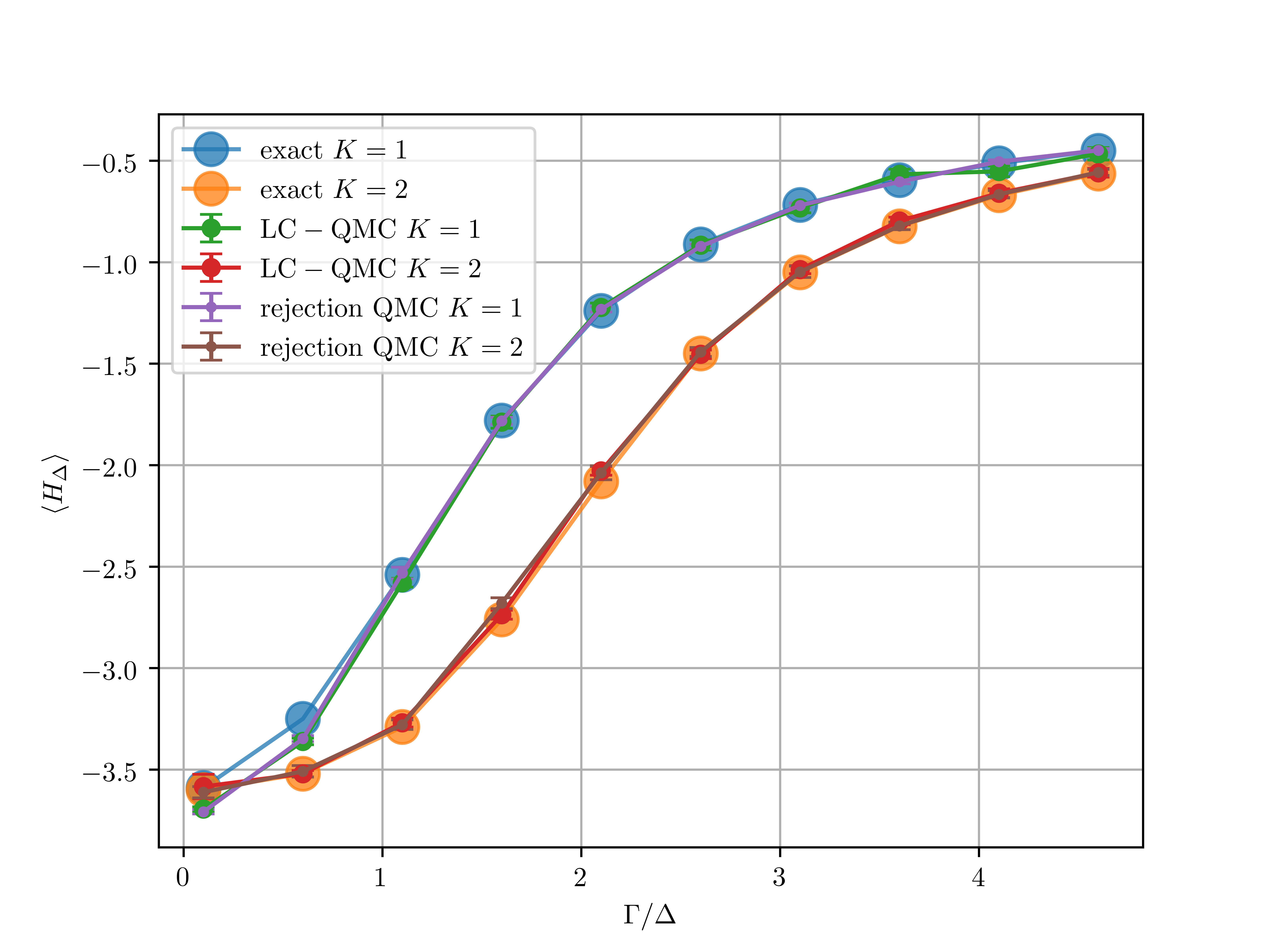}
    \includegraphics[width=0.98\columnwidth]{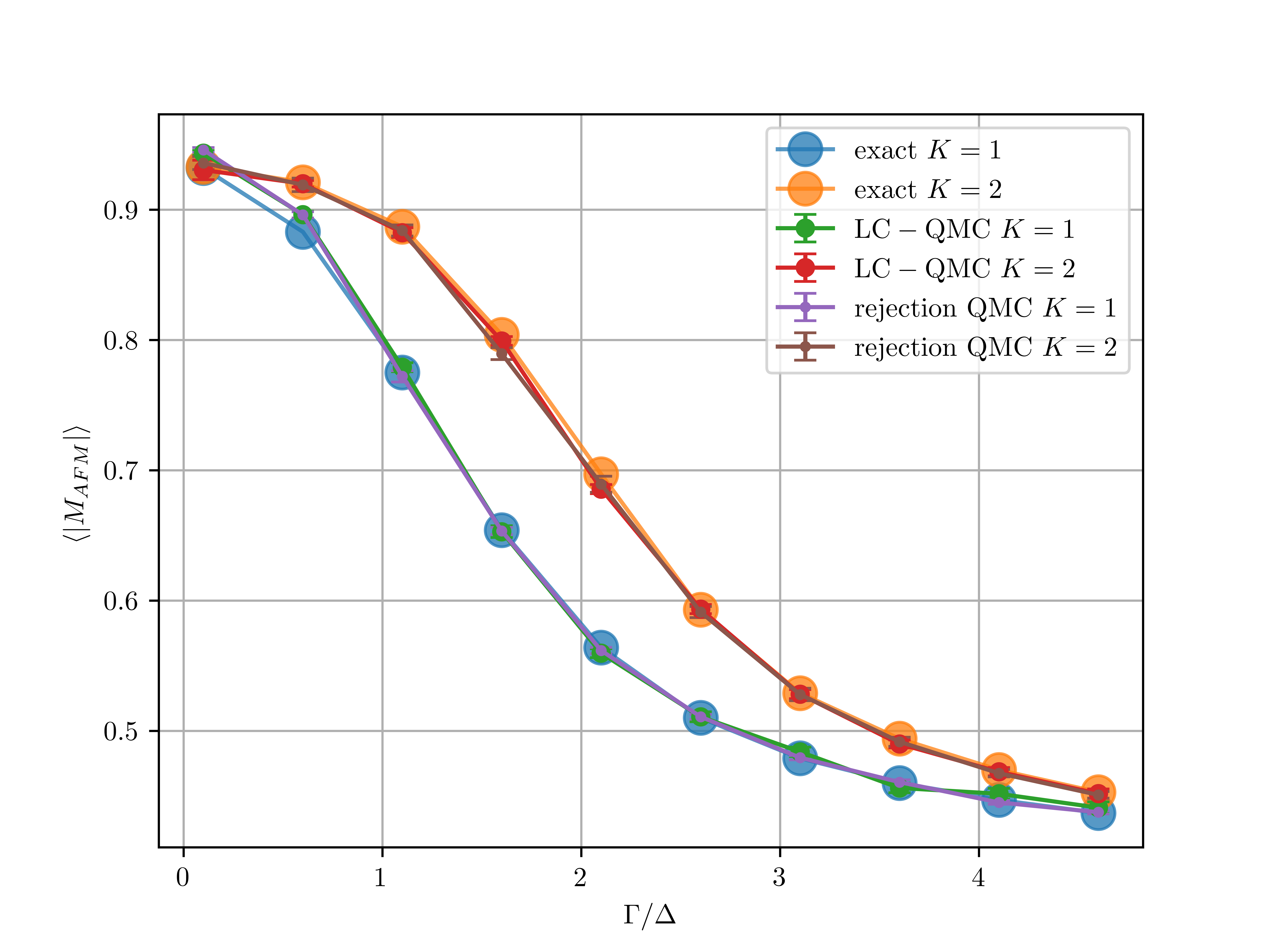}
    \caption{\textbf{Comparison of exact statistics, rejection based QMC, and LC-QMC}. Here we show, for a small exactly solvable system, $L=2$, that the two QMC algorithms agree with the exact result (computed by evaluating the matrix exponential $\exp(-\beta H)$ directly).
     (Top) Average (diagonal) energy $\langle H_\Delta \rangle$. (Bottom) Magnetization.  Each data point in the QMC codes computed using $2^{14}$ sweeps and 75 time-slices. In both plots, $\Delta \beta = 1$.}
    \label{fig:constrained_qmc}
\end{figure}

\begin{figure}
    \centering
    \includegraphics[width=0.98\columnwidth]{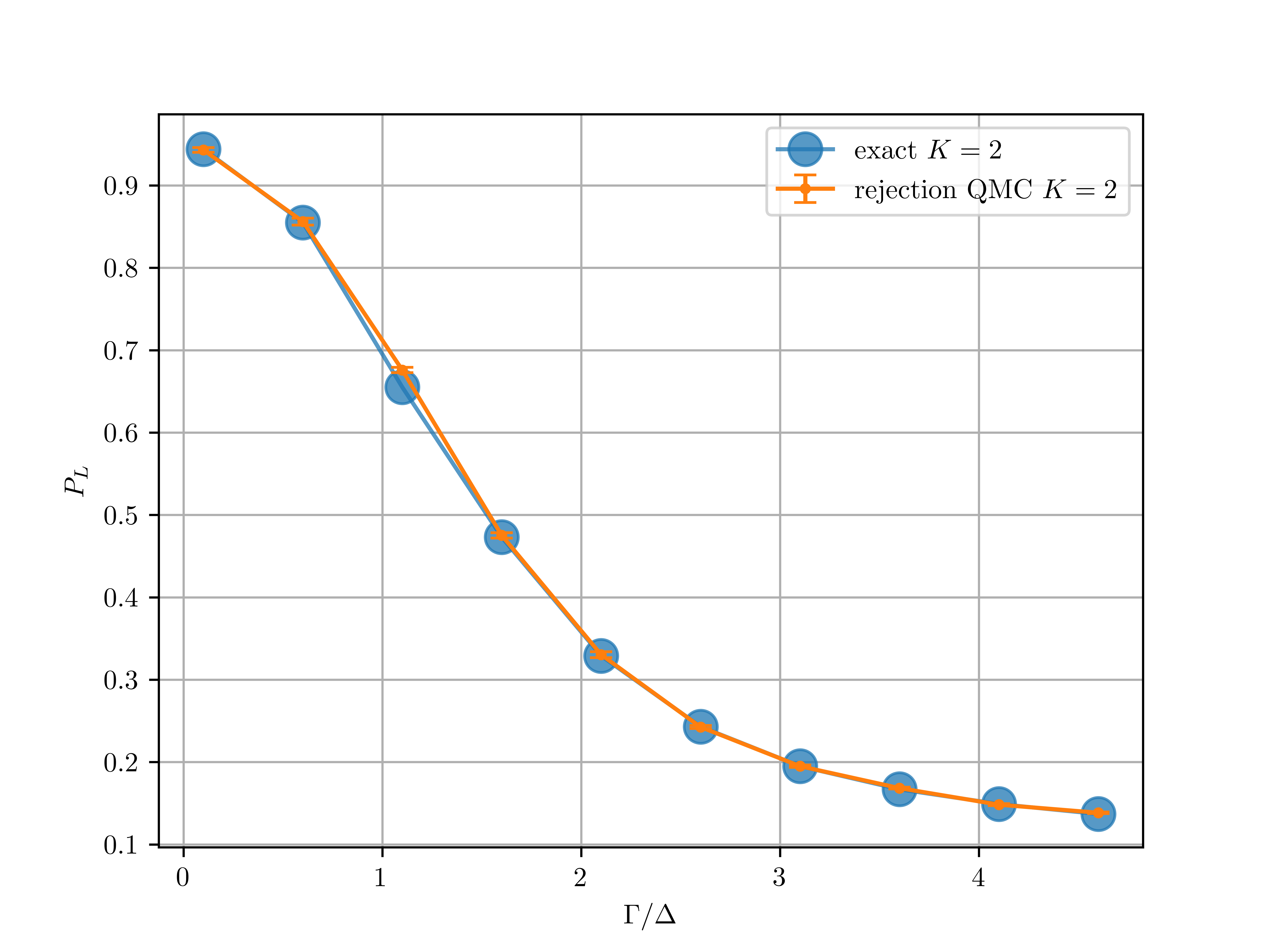}
    \caption{\textbf{Comparison of exact statistics with rejection based QMC for computation of $P_L$}. Here we show, for a small exactly solvable system, $L=2, K=2$, that the QMC rejection based algorithm agree with the exact result of logical probability $P_L$ (computed by evaluating the matrix exponential $\exp(-\beta H)$ directly).
      Each data point in the QMC code computed using $2^{14}$ sweeps and 75 time-slices, with $\Delta \beta = 1$.}
    \label{fig:PL_sample_vs_exact}
\end{figure}

In Fig.~\ref{fig:constrained_qmc} we see how LC-QMC compares to the exact calculation for a small system.
We also compare this to a Rejection-based QMC code where we do not constrain the $\tau=0$ slice, but simply reject any sample which is not a logical one (also see Fig.~\ref{fig:PL_sample_vs_exact}). As explained in the main text, the latter becomes very inefficient, even at modest embedding sizes.

\section{Simulation parameters \label{sect:sim_params}}
\begin{figure}[t!]
    \centering
    \includegraphics[width=0.98\columnwidth]{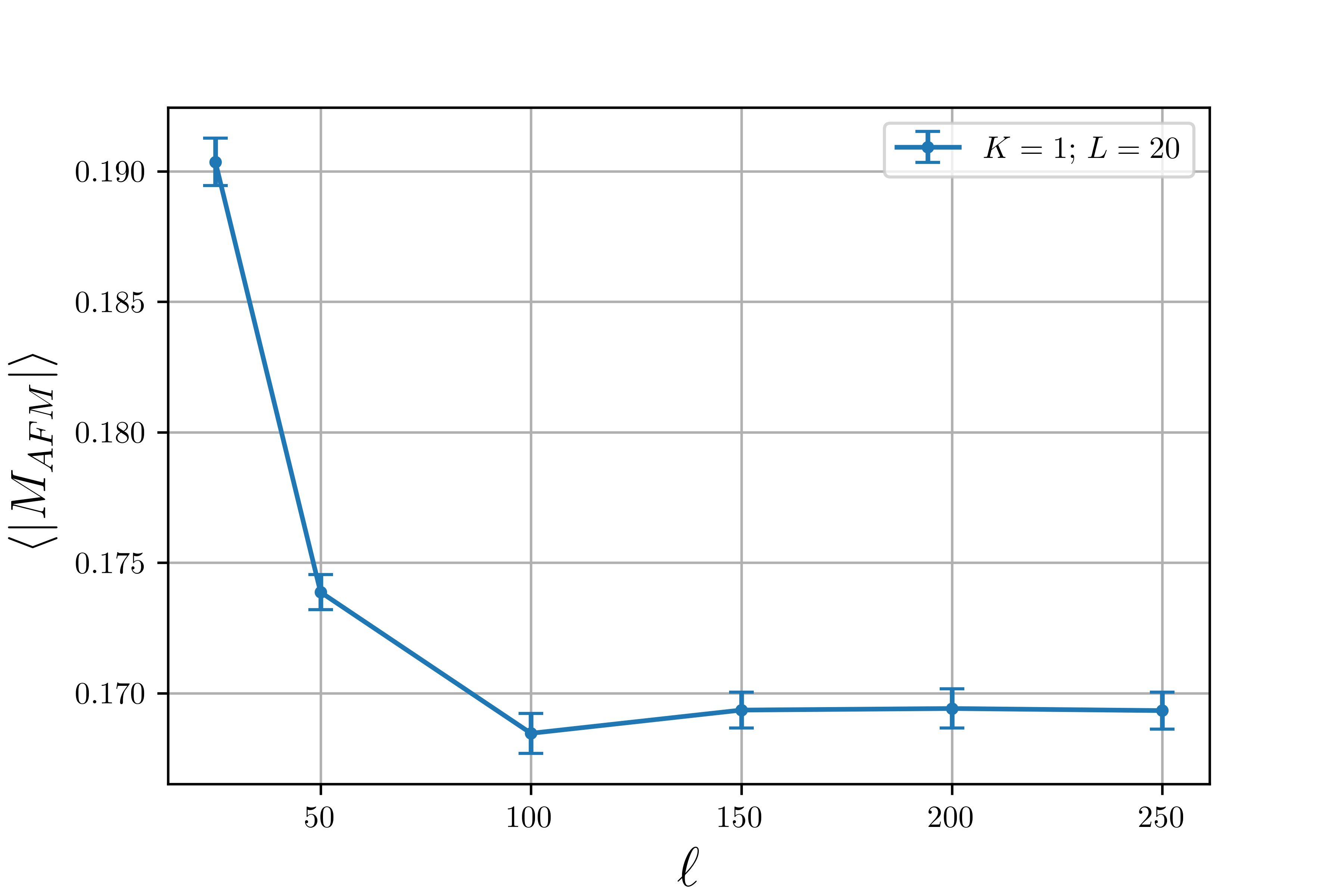}
    \includegraphics[width=0.98\columnwidth]{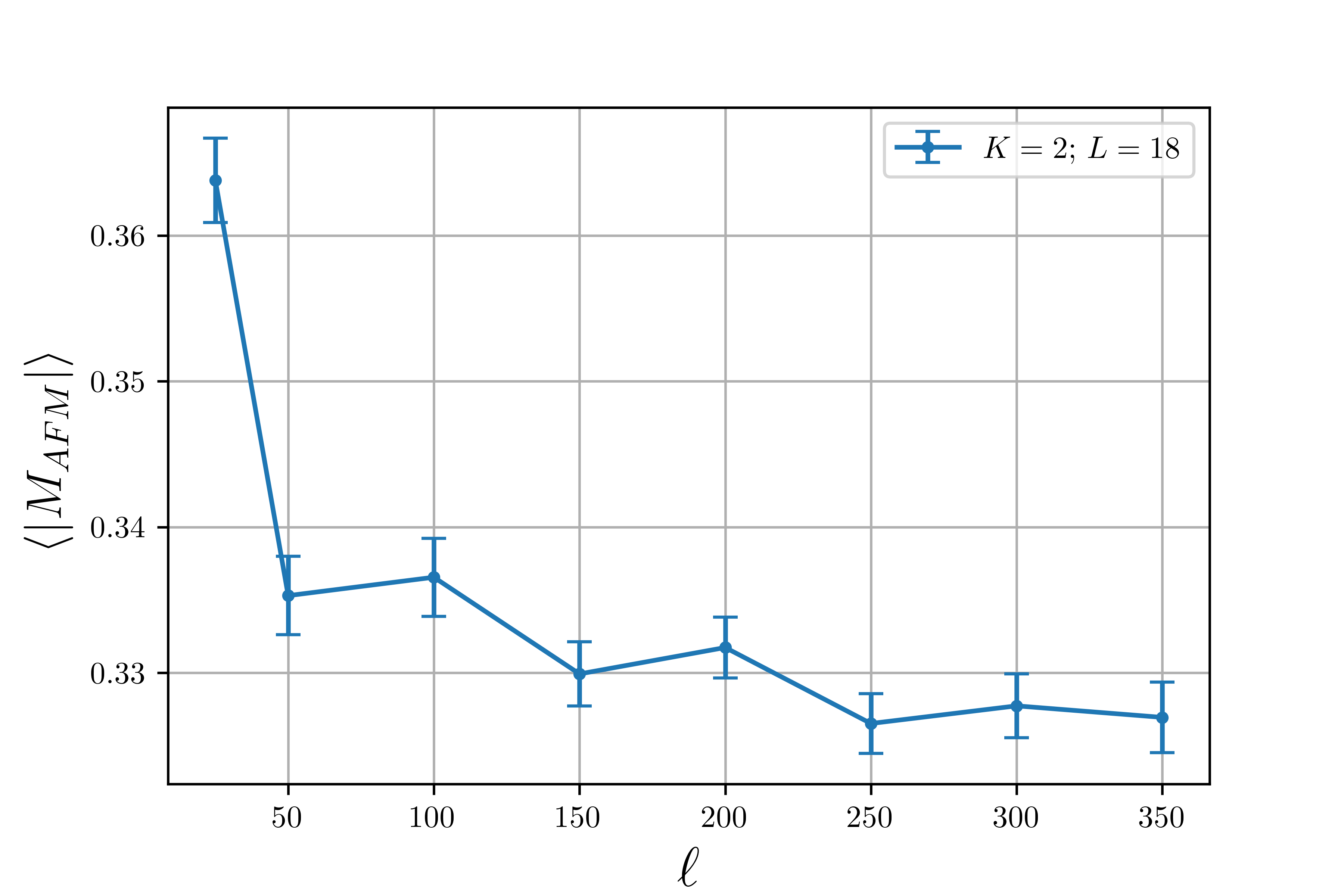}
    \caption{\textbf{Simulation convergence with number of time-slices.} 
    Plotted, for two embedding sizes convergence of the order parameter with number of imaginary time slices $\ell$, for $\Delta \beta=1.645$, and $\Gamma / \Delta = 2.95$ (typical parameters studied in the main text). Shown is data for $L=20, K=1$ (top) and $L=18, K=2$ (bottom) though we do a similar analysis for each system size we studied, in order to pick an appropriate $\ell$ value.
    We see for these sizes, $\ell \gtrsim 150$ suffices for $K=1$, and $\ell \gtrsim 250$ for $K=2$. Each data point is from a single QMC run of $2^{19}$ sweeps, with error bars representing the standard error as discussed in Appendix \ref{sect:sim_params}.}
    \label{fig:tau_converge}
\end{figure}

In Fig.~\ref{fig:tau_converge} we study the convergence of simulation measurements with the number of imaginary time-slices $\ell$ in the LC-QMC. We find $\ell \approx 150$ suffices for $K=1$ and $\ell = 250$ for $K=2$. For larger sizes we do a similar analysis to select the appropriate $\ell$.

\begin{figure}[H]
    \centering
    \includegraphics[width=0.98\columnwidth]{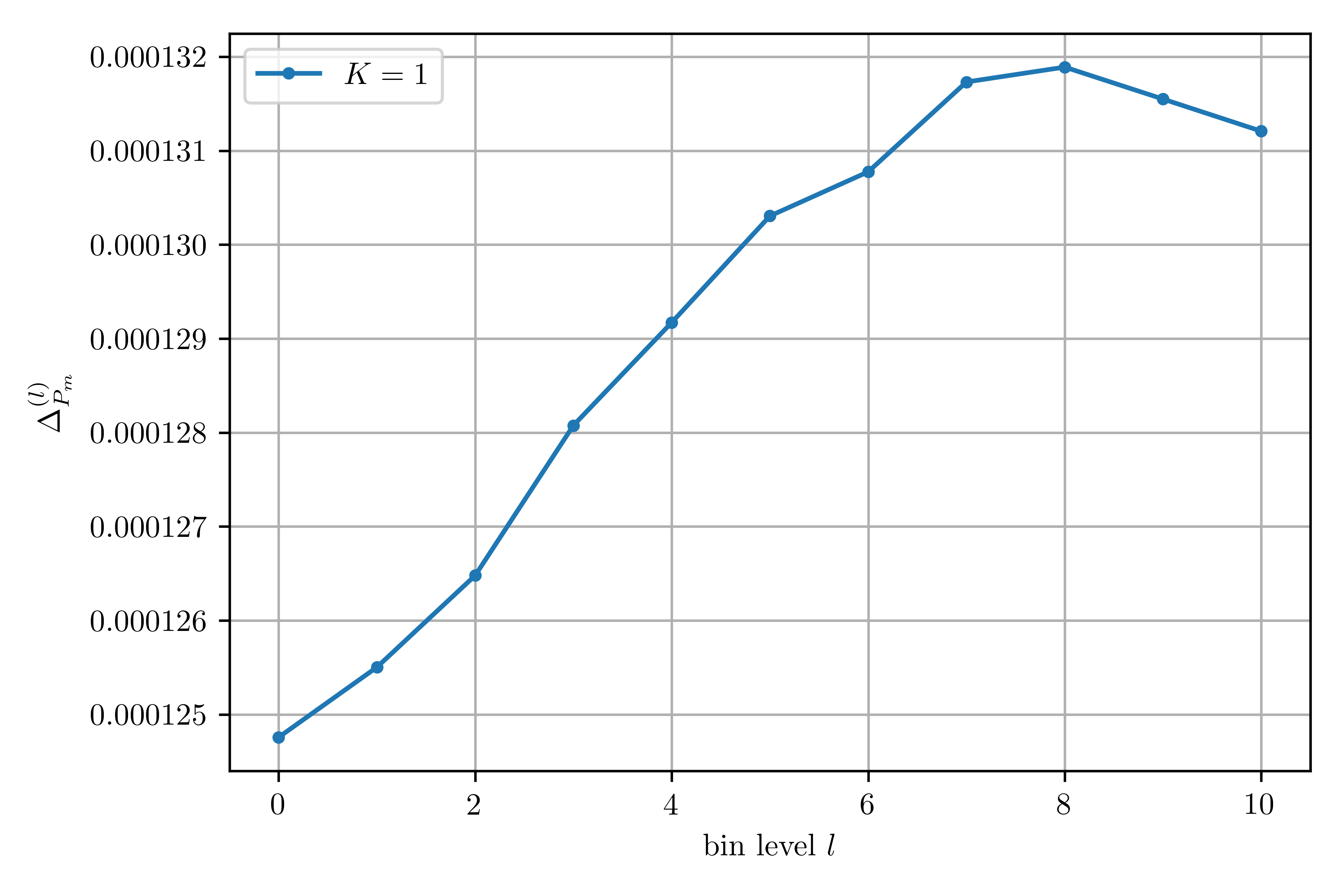}
    \includegraphics[width=0.98\columnwidth]{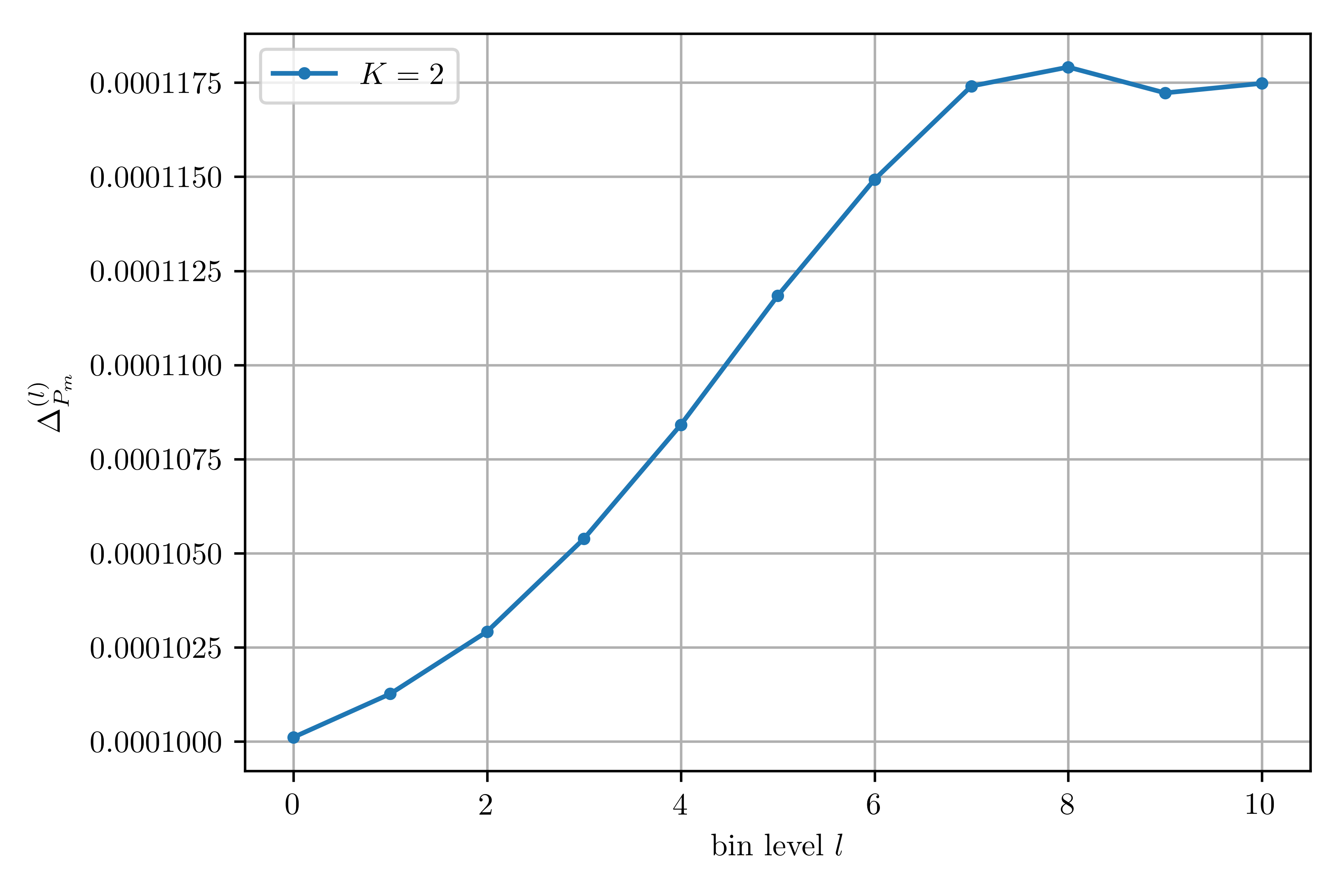}
    \caption{\textbf{Error convergence for order parameter}. For $K=1$ (top) and $K=2$ (bottom), with $L=16$, $\Delta \beta=1.645$, $\Gamma/\Delta = 3.05$, we show convergence of the uncertainty $\Delta_{P_m}$ of the estimate of the probability of order parameter $m=M_{AFM}$, at the peak of the distribution, computed as in Ref.~\cite{mc-errors}. This shows bins of size around $2^8$ ($l=8$) suffice for taking statistics (\ie bins are uncorrelated when at least this size). The error bars we report in the main text is the converged value (\eg $\sim 0.000132$ in the top figure, and $\sim 0.0001175$ bottom). We used $2^{20}$ sweeps to generate these plots, with $\ell=150$ time-slices.}
    \label{fig:hist_errors}
\end{figure}

In order to estimate errors from a single Monte-Carlo simulation, we use a binning analysis, as detailed in Ref.~\cite{mc-errors} (Sect. IVD). In particular, for a single MC run, taking $2^k$ measurement samples of some statistic $A$, we can extract the averages $\langle A \rangle_{i}$ over bins $i$ of size $2^l$, where $l=0,1,\dots$ ($l<k$). Once the bin size is large enough, statistics from different bins are expected to be independent, from which one can compute the standard error.
In Fig.~\ref{fig:hist_errors} we demonstrate that for a typical sized problem studied here, bin sizes of around $2^8$ allows for samples of the order parameter to be decorrelated. When reporting error bars of this type, we report the converged value of the uncertainty (as in Fig.~\ref{fig:hist_errors}).

In some of our simulations, instead of performing a single long QMC run, we perform many independent runs and compute errors bars over these samples instead, using the standard error of the mean for independent samples.

\clearpage

\end{document}